\documentclass[journal]{IEEEtran}
\usepackage[dvips]{graphicx}
\usepackage[cmex10]{amsmath}

\usepackage{mathrsfs}

\usepackage{comment}
\usepackage{bm}
\usepackage{array}

\usepackage{enumitem}
\usepackage{amsmath}
\usepackage{amscd}
\usepackage{amssymb}
\usepackage{latexsym}
\usepackage{epsfig}
\usepackage{cite}
\usepackage{setspace}
\usepackage{multirow}
\usepackage[english]{babel}
\usepackage[table]{xcolor}
\usepackage[margin=2.0cm]{geometry}
\newtheorem{lem}{Lemma}
\newtheorem{ther}{Theorem}
\newtheorem{deft}{Definition}

\newtheorem{Proposition}{Proposition}
\newtheorem{Corollary}{Corollary}
\usepackage{mwe}
\usepackage{algorithm,algcompatible,amsmath}
\algnewcommand\STEP{\item[\textbf{Steps:}]}%

\ifCLASSINFOpdf

\else

\fi

\usepackage[tight,footnotesize]{subfigure}

\begin{document}
\title{Coprime Sensing via Chinese Remaindering over Quadratic Fields, Part I: Array Designs}
\author{Conghui~Li,~
        Lu~Gan,~\IEEEmembership{Senior Member,~IEEE,}
        and~Cong~Ling,~\IEEEmembership{Member,~IEEE}
\thanks{This paper was presented in part at International Conference on Sampling Theory and Applications (SAMPTA), Tallinn, Estonia, July 3-7, 2017.}
\thanks{Conghui Li and Cong Ling are with the Department of Electrical and Electronic
Engineering, Imperial College London, London, SW7 2AZ, U.K. (e-mail:
conghui.li15@imperial.ac.uk; cling@ieee.org).}
\thanks{Lu Gan is with the Department of Electronic and Computer Engineering, Brunel University London, London, UB8 3PH, U.K. (e-mail: lu.gan@brunel.ac.uk).}
}

\markboth{ }%
{Shell \MakeLowercase{\textit{et al.}}: 2-D DOA Estimation by Chinese Remaindering over Rings}

\maketitle

\begin{abstract}
A coprime antenna array consists of two or more sparse subarrays featuring enhanced degrees of freedom (DOF) and reduced mutual coupling. This paper introduces a new class of planar coprime arrays, based on the theory of ideal lattices. In quadratic number fields, a splitting prime $p$ can be decomposed into the product of two distinct prime ideals, which give rise to the two sparse subarrays. Their virtual difference coarray enjoys a quadratic gain in DOF, thanks to the generalized Chinese Remainder Theorem (CRT). To enlarge the contiguous aperture of the coarray, we present hole-free symmetric CRT arrays with simple closed-form expressions. The ring of Gaussian integers and the ring of Eisenstein integers are considered as examples to demonstrate the procedure of designing coprime arrays. With Eisenstein integers, our design yields a difference coarray that is a subset of the hexagonal lattice, offering a significant gain in DOF over the rectangular lattice, given the same physical areas. 
Maximization of CRT arrays in the aspect of essentialness and the superior performance in the context of angle estimation will be presented in the companioning Part II.
\nocite{li2017coprime}
\end{abstract}

\begin{IEEEkeywords}
Array processing, Chinese Remainder Theorem, ideal lattices, sparse arrays.
\end{IEEEkeywords}

\IEEEpeerreviewmaketitle

\section{Introduction}

\IEEEPARstart{A} \normalfont N antenna array is a set of antennas placed in a certain configuration to transmit and/or receive signals. Earlier studies were based on uniform linear arrays (ULAs), uniform rectangular arrays (URAs) and uniform circular arrays (UCAs) by which only a limited number of sources were detected \cite{schmidt1986multiple, roy1989esprit}.

Previous studies have investigated sparse arrays with $O(N)$ physical sensors for one-dimensional (1D) source estimation such as minimum redundancy arrays (MRAs) \cite{chen2008minimum}, nested arrays \cite{pal2010nested}, super nested arrays \cite{liu2016super}, coprime arrays \cite{vaidyanathan2011sparse} and coprime arrays with displaced subarrays \cite{Qin15coprime}, which offer $O(N^2)$ DOF by exploiting the concept of the virtual coarray. Here the DOF of an array is defined as the number of uncorrelated sources that can be identified by the receiver array. Such arrays can identify more sources than the number of sensors because of the enhanced apertures of coarrays. Particularly, among these arrays, super nested arrays and coprime arrays along with their derivatives significantly alleviate the mutual coupling effect among antennas thanks to the sparse geometries. 

In two-dimensional (2D) space, lattices have been well studied in the application of array processing where physical antenna subarrays can be placed on lattices. The virtual coarray is defined as a set of all difference vectors between subarrays.  
In an example of an antenna system with two subarrays, the cross-difference coarray is defined by
\begin{equation*}
 \mathcal{D} = \{ \mathbf{d} :  \mathbf{d} = \mathbf{G}_1\mathbf{x}_2 - \mathbf{G}_2\mathbf{x}_1 \},
\end{equation*}
where $\mathbf{G_1}\mathbf{x_2}$ and $\mathbf{G_2}\mathbf{x_1}$ represent receiver sensor locations; $\mathbf{G}_1$ and $\mathbf{G}_2$ are generator matrices of the subarrays; and $\mathbf{x}_1$ and $\mathbf{x}_2$ are integer vectors \cite{vaidyanathan2011theory,pal2012nested}. By selecting $\mathbf{G}_1$ and $\mathbf{G}_2$ to be commuting and left coprime integer matrices (i.e., there exist two integer matrices $\mathbf{M}$ and $\mathbf{N}$, such that $\mathbf{G_1M}+\mathbf{G_2N=I}$, where $\mathbf{I}$ is the identity matrix), a difference coarray can be obtained, whereby the DOF surges to $O(|\det{(\mathbf{G}_1\mathbf{G}_2)}|)$ with $|\det (\mathbf{G}_1)| + |\det (\mathbf{G}_2)|$ sensors. A method based on Smith Form Decomposition was outlined in \cite{vaidyanathan2011theory} to guarantee the coprimality of $\mathbf{G}_1$ and $\mathbf{G}_2$, whereas in \cite{pal2012nested}, the two generator matrices satisfy the relation $\mathbf{G}_1=\mathbf{G}_2\mathbf{P}$ where $\mathbf{P}$ is a 2-by-2 integer matrix.
More recently, \cite{shi2017sparsity} derived a novel algorithm from the view of the sum-difference coarray where the coprimality of $\mathbf{G}_1$ and $\mathbf{G}_2$ was guaranteed by extending two orthogonal 1D coprime arrays. 
Examples of non-lattice based sparse arrays include  
\cite{liu2017hourglass}, which redistributed the open-box array \cite{doi} to reduce the mutual coupling effect and possess the hole-free property.
Nevertheless, the coarray has been restricted to a subset of $\mathbb{Z}$ or $\mathbb{Z}^2$ in previous studies \cite{pal2012nested,chen2008minimum,pal2010nested,vaidyanathan2011sparse,vaidyanathan2011theory,liu2017hourglass,liu2016super,Qin15coprime, shi2017sparsity}.

This paper along with its companion paper further completes the investigation of coprime array design by means of the Chinese remainder theorem (CRT) over quadratic fields. The classical CRT allows the reconstruction of a rational integer from its remainders by pairwise coprime divisors. A crucial consequence of this theorem is that it can be extended to a general form in ring theory, which allows the computation of algebraic integers by rephrasing the classical CRT in terms of ideals and rings \cite{marcus1977number}. 
As a result, the coprime arrays introduced in \cite{vaidyanathan2011sparse} and \cite{vaidyanathan2011theory} can be interpreted as cases of CRT over $\mathbb{Z}$ and over $\mathbb{Z}^2$ respectively. 
Herein, we relate pairwise coprime algebraic integers to multi-dimensional lattices in Euclidean space through canonical embedding. Specifically, we apply CRT over \emph{rings of algebraic integers} to construct coprime subarrays, which are subsets of \emph{ideal lattices} arising from the prime decomposition. However, the conditions pertaining to the coprimality of algebraic integers and ideal lattices are non-trivial.

This paper shows the connection between coprime algebraic integers and their corresponding lattices that are obtained by embeddings and represented by integer matrices. In general, the coprimality of integer matrices is defined in matrix rings \cite{vidyasagar2011control, Introductiontonumbertheory}. The class of integer matrices obtained from algebraic integers in this work may be seen as special matrix rings. Principal advantages of relating algebraic integers with matrices include commutativity, simplified expressions and the potential to exploit the nice properties of algebraic integers (e.g., the convenience to check coprimality).
For instance, the coprimality issues of some classes of matrices such as adjugate pairs and skew circulant pairs \cite{vaidyanathan2011general,pal2011coprimality} can be addressed as special cases of algebraic conjugate integers.  
Examples of algebraic integers in \emph{quadratic fields} including \emph{ring of Gaussian integers} and \emph{ring of Eisenstein integers} are studied in this paper.

An important advantage of \emph{Eisenstein integers} is that the difference coarray becomes a subset of the hexagonal lattice $A_2$. 
It is well known that $A_2$ is the optimum lattice for sphere packing in two-dimensional space \cite{BK:Conway98}. Numerical analysis reveals that the optimum packing density results in a $15.5\%$ gain in DOF for a fixed physical area of the array. Due to this reason, the hexagonal geometry is currently used in the design of some phased-array antennas \cite{tian1998doa,van2004optimum}. This paper together with its accompanying paper puts forward the application of hexagonal lattices
and hence provides the potential to decrease the physical array aperture without sacrificing DOF.

The main contribution of this paper along with its accompanying paper is that they further develop the design methods of 2D coprime arrays, which brings a new class of 2D array configurations, namely \emph{CRT arrays}. Such arrays can provide enhanced DOF, sparse geometry, and hole-free coarrays. By lattice representations, the configurations of the proposed arrays along with their virtual coarrays enjoy simple closed-form expressions.
This paper addresses the issues relating to geometry and the generation of CRT arrays including the mapping between number fields and lattices, coprimality issues pertaining quadratic integers and embedded matrices, and lattice representations of CRT arrays along with their coarrays, whereas the accompanying part II employs CRT arrays to propose practical algorithms for angle estimations in both active and passive sensing scenarios.   
Other potential applications of Chinese remaindering over quadratic fields include sparse 2D discrete fourier transform \cite{guessoum1986fast}, the radar measurements on multiple targets \cite{dahl2015comparison, dahl2016mimo}, filter banks\cite{vaidyanathan2011theory}, imaging \cite{wu2015high,hoctor1990unifying} and direction finding problems using compressive sensing \cite{qin2017doa}.

The rest of the paper is organized as follows. Before constructing coprime lattices from prime ideals in Section \ref{Ideals in quadratic fields and construction of the ideal lattices}, the concepts of quadratic fields and their rings of integers are briefly reviewed in Section \ref{Quadratic fields and Prime Elements} along with algebraic lattices. 
Based on CRT, Section \ref{Design of CRT-based Sparse Arrays} proposes a new class of coprime arrays allocated on coprime lattices and derives closed-form expressions of CRT array geometries, which are inherent in any \emph{rings of algebraic integers}. Section \ref{Hole-free CRT-based Array} extends CRT to hole-free symmetric CRT whereby the parameter identifiability is enhanced, after which examples are provided for $\mathbb{Z}^2$ and $A_2$ as special cases of quadratic integers. Section \ref{Conclusion} concludes the paper.

\emph{Notations:} Bold font lowercase letters (e.g., $\mathbf{x}_1$), bold font uppercase letters (e.g., $\mathbf{G}$), fraktur font letters (e.g., $\mathfrak{p}_1$) and calligraphy font alphabets (e.g., $\mathcal{D}$) denote vectors, matrices, principal  ideals and sets respectively. $\mathbb{Z}$ and $\mathbb{Q}$ denote rational integers $\{ \cdots -1, 0, 1 \cdots \} $ and rational numbers $\{ \frac{a}{b} \; | \; a,b \in \mathbb{Z}, \; b\neq 0 \}$ respectively. 
$\mathbb{R}^F$ denotes the $F$-dimensional Euclidean space. $\operatorname{Re}(m)$ and $\operatorname{Im}(m)$ represent the real and imaginary parts of a complex number $m$ respectively.
$\text{N}(m)=m\hat{m}$ denotes the norm of $m$ where $\hat{m}$ is the algebraic conjugate of $m$ (Section \ref{Quadratic Field and Its Ring of Integers}). For example in the ring of Gaussian integers, with $m=3+2i$, it can be readily verified that $\text{N}( m )=m\hat{m}=13$.

\section{Review of Quadratic Fields and Algebraic Lattices}\label{Quadratic fields and Prime Elements}

Let us first briefly review some definitions and preliminary results related to \emph{quadratic field} along with its \emph{ring of integers}; and based on \emph{algebraic integers}, the construction of \emph{algebraic lattices}, on which sensor arrays can be allocated. \cite{marcus1977number,vaidyanathan2011sparse} 

\subsection{Quadratic Field and Its Ring of Integers}\label{Quadratic Field and Its Ring of Integers}
A quadratic field $K$ is a field extension of degree 2 over rational numbers $\mathbb{Q}$, i.e., it is a $\mathbb{Q}$-vector space of dimension two. Note that $ \mathbb{Q} \subseteq K$. For instance, $\sqrt{-1}$ is not an element in $\mathbb{Q}$ but it is an element in the field extension of $\mathbb{Q}$. In order to be a field, this new field extended from $\mathbb{Q}$ must contain $\mathbb{Q}$ and all the powers and multiples of $\sqrt{-1}$. In other words, $\mathbb{Q}$ is extended into a new vector space over $\mathbb{Q}$, which is generated by the powers of $\sqrt{-1}$. Let $i \triangleq \sqrt{-1}$ and $\mathbb{Q}(i)$ denote this field extension. Every element $m \in \mathbb{Q}(i)$ can be expressed as $m=m_1+m_2i$, $m_1, m_2 \in \mathbb{Q}$, i.e., $\{ 1, i\}$ is the basis of $\mathbb{Q}(i)$. In this case, an \emph{algebraic integer} takes the form of $m_1+m_2i$  where $m_1,m_2, \in \mathbb{Z}$. The ring of integers of $\mathbb{Q}(i)$ is the set of all algebraic integers in $\mathbb{Q}(i)$, which can be represented by $\mathbb{Z}[i]=\{ m_1+m_2i, m_1,m_2, \in \mathbb{Z} \}$.  

In general, a quadratic field is denoted by $K=\mathbb{Q}(\sqrt{D})$, where $D$ is a square-free rational integer. Note that if $D$ is a perfect square, $K=\mathbb{Q}$. The ring of integers is often denoted as $\mathcal{O}_K$, which is a set that contains all algebraic integers in $K$. In quadratic fields, algebraic integers are also known as \emph{quadratic integers} which are roots of quadratic polynomials with coefficients in $\mathbb{Z}$. The minimal polynomial denoted as $f(X)$ of $\mathcal{O}_K$ can be expressed as:
\begin{equation}\label{fx}
f(X)=\left\{
       \begin{array}{ll}
       X^2 - D , & \hbox{if $D \not\equiv 1 \pmod{4}$ ;} \\
       X^2 - X + \frac{1-D}{4}, & \hbox{if $D \equiv 1 \pmod{4}$, }
       \end{array}
       \right.
\end{equation}
or alternatively, 
\begin{equation}\label{fx2}
f(X)=X^2+BX+C,    
\end{equation}
where $B=0$ and $C=-D$ when $D \not\equiv 1 \pmod{4}$, and $B=-1$ and $C=\frac{1-D}{4}$ when $D \equiv 1 \pmod{4}$. The proof of $f(X)$ can be found in \cite{marcus1977number}. Let $q$ and $\hat{q}$ denote the two roots of $f(X)$ respectively. With the notations above, it can be easily calculated that 
\begin{equation}\label{q}
     q = -\frac{1}{2}B + \frac{1}{2}\sqrt{B^2-4C}, \: \, \text{and}
\end{equation}
\begin{equation}\label{hatq}
     \hat{q} = -\frac{1}{2}B - \frac{1}{2}\sqrt{B^2-4C}
\end{equation}
Here $\{ 1, q \}$ and $\{ 1, \hat{q} \}$ are called the \emph{integral bases} of $\mathbb{Q}(\sqrt{D})$ \cite{marcus1977number}, i.e., every element in $\mathbb{Q}(\sqrt{D})$ can be written as $m_1+m_2q$ corresponding to the former basis or as $m_1+m_2\hat{q}$ corresponding to the latter with $m_1, m_2 \in \mathbb{Q}$.
Accordingly, every element in its ring of integers can be formed by $m=m_1+m_2q$ or $\hat{m}=m_1+m_2\hat{q}$ with $m_1, m_2 \in \mathbb{Z}$.
Here $m_1+m_2q$ and $m_1+m_2\hat{q}$ are called \emph{algebraic conjugates} of each other, which can be viewed as a generalization of the complex conjugation. From (\ref{q}) and (\ref{hatq}), it can be verified that with $B^2-4C<0$ ($D<0$), the two conjugations are identical to each other.

With the knowledge of the integral basis, the ring of integers $\mathcal{O}_K$ can be denoted as $\mathbb{Z}[q]$, and $m=m_1+m_2q$ is called a quadratic integer of $\mathbb{Z}[q]$ for any $m_1$ and $m_2$ in $\mathbb{Z}$, which generalizes rational integers in $\mathbb{Z}$ to quadratic fields. Note that $q \neq \hat{q}$ since $D$ is square-free, whereas $\mathbb{Z}[q]$ and $\mathbb{Z}[\hat{q}]$ represent the same ring of integers as $\hat{q} = -B -q$. Henceforth, we will use $\mathbb{Z}[q]$ as the notation of the ring of integers of $\mathbb{Q}(\sqrt{D})$. 

When $D=-1$, for example, $f(X) = X^2+1$ whose roots are $q=i$ and $\hat{q}=-i$. The ring of integers of $\mathbb{Q}(i)$ denoted by $\mathcal{O}_K=\mathbb{Z}[i]$ is also known as the ring of \emph{Gaussian integers}. In this case, $\{ 1, i\}$ is an integral basis of $\mathbb{Z}[i]$, since $ -1 \equiv 3 \pmod{4}$. Therefore, every element in $\mathbb{Z}[i]$ can be uniquely expressed as $m_1+m_2i$ with $m_1, m_2 \in \mathbb{Z}$. The algebraic conjugation of $m$ is $\hat{m}=m_1+m_2\hat{q}=m_1-m_2i$ which is also the complex conjugation of $m$. 
Another example that will be used in this paper for illustrative purposes is the ring of \emph{Eisenstein integers} with $D=-3$. In this case, $\mathcal{O}_K=\mathbb{Z}[\omega]$ with $\{1, \omega \}$ as its integral basis where $\omega=e^{i\pi/3}=\frac{1}{2}+i\frac{\sqrt{3}}{2}$ since $D \equiv 1 \pmod{4}$. An arbitrary element in $\mathbb{Z}[\omega]$ can be expressed as $n=n_1+n_2\omega$ with $\hat{n}=n_1+n_2\hat{\omega}$ being its algebraic conjugation where $\hat{\omega}=\frac{1}{2}-i\frac{\sqrt{3}}{2}$.   

\subsection{Construction of Algebraic Lattices}

\begin{deft}\label{voronoi cell}
Given $F$ linearly independent column vectors $\mathbf{g}_1, \mathbf{g}_2,\ldots, \mathbf{g}_F\in \mathbb{R}^F $, an $F$-dimensional lattice $\Lambda$ is defined as the set of integer combinations of the basis vectors, \textit{i.e.},
\begin{eqnarray*}
\Lambda = \left\{ \sum_{k=1}^F{x_k\mathbf{g}_k} : x_k \in \mathbb{Z} \right\}.
\end{eqnarray*}
Accordingly, the generator matrix of the lattice $\Lambda$ is obtained by
\begin{eqnarray*}
\mathbf{G} = [ \mathbf{g}_1 | \mathbf{g}_2 | \cdots | \mathbf{g}_F].
\end{eqnarray*}
\end{deft}
The Voronoi cell of $\Lambda$ is defined by
\begin{equation}\label{Vorinoi regin}
  \mathcal{V}(\Lambda)=\{\mathbf{y}\in \mathbb{R}^F: \|\mathbf{y}\| \leq \|\mathbf{y}-\lambda\|, \forall \lambda \in \Lambda \},  
\end{equation}
where ties are broken in a systematic manner. 

There are various ways to construct lattices, for instance, from codes and groups. In this paper, lattices and their ideals are obtained from rings of quadratic integers $\mathbb{Z}[q]$ via the canonical embedding. 

In general, the canonical embedding builds a bridge between lattices and rings of algebraic integers as it establishes a bijective mapping between the elements in an algebraic number field of degree $F$ and the vectors of the $F$-dimensional Euclidean space. In other words, the canonical embedding $\sigma$ sends an algebraic integer $m$ to a lattice point $\mathbf{m}=\sigma(m)$ in Euclidean space where $\mathbf{m}$ is an $F$-by-1 vector. The canonical embedding of any algebraic number field of degree $F$ is given in \cite[Definition 5.15]{oggier2004algebraic}. 

Herein we consider quadratic fields where $F=2$. Then the embeddings of $\mathbb{Q}(\sqrt{D})$ are simply given by
\begin{equation*}
    \sigma_1(\sqrt{D})=\sqrt{D}, \quad \text{and} \quad \sigma_2(\sqrt{D})=-\sqrt{D}.
\end{equation*}
The canonical embedding $\sigma$ is a geometrical representation of $\mathbb{Q}(\sqrt{D})$ that maps $m \in \mathbb{Q}(\sqrt{D})$ to a vector of 2D Euclidean space, i.e., $\sigma(m) = (\sigma_1(m), \sigma_2(m))^T \in \mathbb{R}^2 $. 
For example, the embeddings of an arbitrary element $m = m_1 + \sqrt{2}m_2 \in \mathbb{Q}(\sqrt{2})$ are given by $\sigma_1(m)=m_1+\sqrt{2}m_2$ and $\sigma_2(m)=m_1-\sqrt{2}m_2$. Therefore, an algebraic lattice can be constructed by embeddings as follows: 

Given $\{1, q \}$ as an integral basis of $\mathbb{Q}(\sqrt{D})$, a 2D \emph{algebraic lattice} $\Lambda = \Lambda(\mathcal{O}_K) = \sigma(\mathcal{O}_K)$ is a lattice whose generator matrix is explicitly given by
\begin{equation}\label{Greal}
\mathbf{G}=\left(
    \begin{array}{cc}
      1 &  \sigma_1(q)\\
      1 & \sigma_2(q) \\
    \end{array}
  \right) \text{ for } D>0,
\end{equation}
whereas if $D<0$, $\mathbb{Q}(\sqrt{D})$ is also known as an imaginary quadratic field where the canonical embedding is further simplified and can be formulated by $\sigma(m) = (\operatorname{Re}(m) , \operatorname{Im}(m))^T$. Hence the corresponding generator matrix is computed by stacking the real and imaginary parts of $1$ and $q$:
\begin{equation}\label{Gimaginary}
\mathbf{G}=\left(
    \begin{array}{cc}
      1 &  \operatorname{Re}(q)\\
      0 & \operatorname{Im}(q) \\
    \end{array}
  \right) \text{ for } D<0.
\end{equation}
Note that $\mathbf{G} $ is a non-singular matrix whose absolute determinant equals to the fundamental volume of its corresponding lattice, i.e.,  $V(\Lambda) = |\det(\mathbf{G})| $ \cite[Theorem 5.8]{oggier2004algebraic}.

\begin{figure}[tb]
  \begin{center}
    \includegraphics[height=2in]{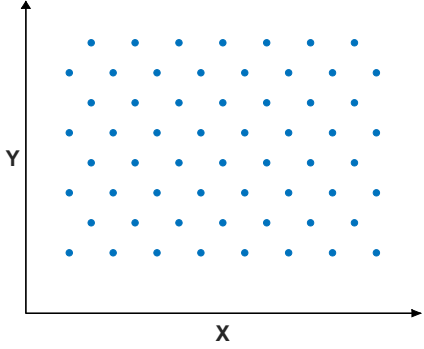}
    \caption{An illustration of $A_2$ lattice.}
      \label{FigEisensteinLattice}
  \end{center}
  \vspace{-1em}
\end{figure}

For example, in the case of $D=-3$ (Eisenstein integers), since the integral basis is $\{1,\omega\}$, the generator matrix of the corresponding algebraic lattice is given by
\begin{equation}\label{Gee}
\mathbf{G}_E=\left(
    \begin{array}{cc}
      1 &  \operatorname{Re}(\omega)\\
      0 & \operatorname{Im}(\omega) \\
    \end{array}
  \right)
  =\left(
    \begin{array}{cc}
      1 &  \frac{1}{2}\\
      0 & \frac{\sqrt{3}}{2} \\
    \end{array}
  \right).
\end{equation}
The lattice constructed from $\mathbf{G}_E$ is shown in Fig. \ref{FigEisensteinLattice}, which is also known as the hexagonal lattice $A_2$ with the densest sphere packing in dimension two. The fundamental volume of $A_2$ is $V(A_2)  =\sqrt{3}/2$ with a minimum distance $1$.

\begin{figure}[tb]
  \begin{center}
    \includegraphics[height=2in]{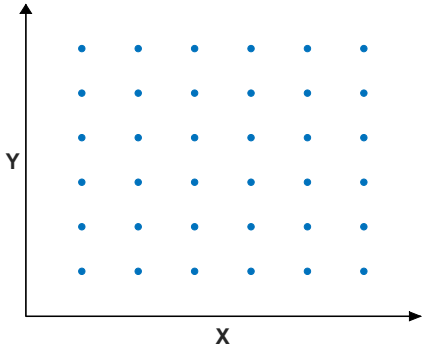}
    \caption{An illustration of $\mathbb{Z}^2$ lattice.}
    \label{FigGuassianLattice}
  \end{center}
  \vspace{-1em}
\end{figure}

Analogously, the ring of Gaussian integers gives rise to the integer lattice $\mathbb{Z}^2$ whose  generator matrix is
\begin{equation}\label{Gg}
\mathbf{G}_G=\left(
    \begin{array}{cc}
      1 &  \operatorname{Re}(i)\\
      0 & \operatorname{Im}(i) \\
    \end{array}
  \right)
  =\left(
    \begin{array}{cc}
      1 &  0\\
      0 & 1 \\
    \end{array}
  \right).
\end{equation}
Fig. \ref{FigGuassianLattice} depicts the configuration of $\mathbb{Z}^2$.

In general, given a number field of degree $F$, its ring of integers $\mathcal{O}_K$ can always construct the algebraic lattice that is expressed by means of the generator matrix.
This construction provides a general and straightforward way of finding pairwise coprime matrices from pairwise coprime algebraic integers, which significantly simplifies the method of matrix factorization from Smith form \cite{lin1994new, vaidyanathan2011theory} and extends integer matrices in Smith form to any matrices that correspond to coprime elements in $\mathcal{O}_K$.

\section{Prime Ideals in quadratic fields and construction of ideal lattices}\label{Ideals in quadratic fields and construction of the ideal lattices}

In the previous section, the construction of 2D algebraic lattices from quadratic fields was briefly reviewed. Similar to 1-D arrays \cite{vaidyanathan2011sparse} where two coprime rational integers in $\mathbb{Z}$ were applied to determine sensor positions, in 2D array design, the quadratic integers in $\mathbb{Z}[q]$ shall be coprime as well, to which Chinese Remainder Theorem applies. Therefore, this section studies prime quadratic integers along with its corresponding prime ideals, from which the algebraic lattices will be constructed.
The computation of prime ideals and the issue of coprimality pertaining algebraic conjugates will be addressed. Examples are provided in Gaussian and Eisenstein integers, which will be exploited to design CRT arrays in the following sections.     

\subsection{Prime Elements in Quadratic Fields}\label{Prime Elements in Quadratic Fields}
 
To distinguish from prime numbers in $\mathbb{Z}$ (e.g., $\pm2, \pm3, \pm5 ,\pm7  \cdots$), a non-zero element $m$ in $\mathbb{Z}[q]$ is a \emph{prime element} if and only if it is not a unit of $\mathbb{Z}[q]$ and whenever $m$ divides a product in $\mathbb{Z}[q]$, it also divides one of the factors. 
Herein, the unit is defined as a quadratic integer $u \in \mathbb{Z}[q]$ with $\text{N}(u)=\pm 1$. In the case of Gaussian integers, there are four units: $\pm 1, \pm i$, and in $\mathbb{Z}[\omega]$, the six units are $\pm 1, (\pm 1 \pm \sqrt{3})/2$.    

For example, 7 is a prime number in $\mathbb{Z}$ but not a prime element in $\mathbb{Z}[\omega]$ since it is reducible, i.e., $7=(1+2\omega)(3-2\omega)$. 
Analogous to the norm of a complex number, with the notations above, the norm of a quadratic integer $m=m_1+m_2q$ in $\mathbb{Z}[q]$ is defined as the product of $m$ and its algebraic conjugate $\hat{m}$, i.e., $\text{N}(m) = m\hat{m}$. As $q$ and $\hat{q}$ are roots of Equation (\ref{fx2}), $q+\hat{q}=-B$ and $q\hat{q}=C$. Thus $\text{N}(m)$ can be derived as follows: 
\begin{equation}\label{normm}
\begin{aligned} 
  & \text{N}(m) = m\hat{m} = (m_1 + m_2 q )(m_1 + m_2 \hat{q}) \\
  & = m_1^2 +m_1m_2(q+\hat{q}) + m_2^2q\hat{q}  
  = m_1^2-Bm_1m_2 +Cm_2^2.  \\
\end{aligned}   
\end{equation}
Since $m_1, m_2, B, C \in \mathbb{Z}$, the norm of a quadratic integer is always in $\mathbb{Z}$. In general, for all $m \in \mathbb{Z}[q]$, it can be verified from \cite[Theorem 1.8]{mollin2011algebraic} that if $\text{N}(m)$ is a prime number in $\mathbb{Z}$, then $m$ is a prime element. 
In the cases of Gaussian and Eisenstein primes, the sufficient and necessary conditions of prime elements can be derived \cite{pollard1998theory}.

A \emph{Gaussian prime} is a prime element in the ring of Gaussian integers of the form $m_1+m_2i$ that satisfies one of the following: 
\begin{itemize}
  \item Both $m_1$ and $m_2$ are nonzero and $\text{N}(m)=m_1^2 + m_2^2$ is a prime number. 
  \item $m_1=0$ and $ m_2\neq 0$ (or $ m_1 \neq 0 $ and $m_2=0$), $m_2$ is a prime number and $| m_2| \equiv 3 $ (mod 4) (or $m_1$ is a prime number and $| m_1 | \equiv 3 $ (mod 4)).
\end{itemize}
In $\mathbb{Z}[\omega]$, an Eisenstein integer of the form $n=n_1+n_2\omega$ is a \emph{Eisenstein prime} if either:
\begin{itemize}
  \item $n_1+n_2\omega$ equals to the product of a unit and a prime number of the form $3a-1$, $a\in \mathbb{Z}$, or
  \item $\text{N}(n)=n_1^2+n_1n_2+n_2^2$ is a prime number.
\end{itemize}
For example, $2+i$ is a Gaussian prime because $1^2+2^2=5$ is a prime number. Likewise,  $2+\sqrt{3}i=1+2\omega$ is a Eisenstein prime since $1^2 + 2 + 2^2 = 7 $ is a prime number. 

\subsection{Prime Factorization into Ideals}

An \emph{ideal} $\mathcal{I}$ in a quadratic field is a subset of $\mathbb{Z}[q]$ such that whenever $x \in \mathcal{I}$ and $m \in \mathbb{Z}[q]$, $mx$ belongs to $\mathcal{I}$. If the ideal is generated by a single element in $\mathbb{Z}[q]$, this ideal is called a \emph{principal ideal}. For instance, $5\mathbb{Z}$ of $\mathbb{Z}$ is a principal ideal whose elements are $ \pm 5, \pm 10 \pm 15 \cdots  $. For simplicity, we can write $5\mathbb{Z} = \langle \,5 \, \rangle$. 
Similar to the prime elements in $\mathbb{Z}[q]$ mentioned in Section \ref{Prime Elements in Quadratic Fields}, a \emph{prime ideal} is an ideal such that $ mn \in \mathcal{I}$ implies $m \in \mathcal{I}$ or $n \in \mathcal{I}$. 
In a \emph{principal  ideal domain} (PID) where every ideal is principal, prime ideals are simply generated by prime elements. All quadratic fields with class number one are PIDs. It has been proved that the PIDs in imaginary quadratic fields are $\mathbb{Q}(\sqrt{D})$, for $D= - 1, -2, -3, -7, -11, -19, -43, -67, -163$, while the full list of PIDs in real quadratic fields is not known yet. Examples include  $D= 2, 3, 5, 6, 7, 11, 13, 14, 17, 19 \cdots$ \cite{neukirch1999algebraic}.  
For simplicity, we only consider PIDs henceforth.
For example,
$ \langle \, 2+\sqrt{3}i \, \rangle $ is a prime ideal of the ring of integers of $\mathbb{Q}(\sqrt{-3})$, namely $\mathbb{Z}[\omega]$, since $\mathbb{Q}(\sqrt{-3})$ is a PID and $2+\sqrt{3}i$ is a prime element (Section \ref{Prime Elements in Quadratic Fields}). The elements in $ \langle \, 2+\sqrt{3}i \, \rangle $ are in the set $ \langle \, 2+\sqrt{3}i \, \rangle = \langle \, 2+\sqrt{3}i \, \rangle \mathbb{Z}[\omega] = \{ (2+\sqrt{3}i)m : m \in  \mathbb{Z}[\omega] \}$.   

Similar to the fundamental theorem of arithmetic to rational integers \cite[Theorem 5.3]{Introductiontonumbertheory}, every non-zero element in a unique factorization domain (UFD) can be written as a product of prime elements. 
More generally, a nonzero ideal $\langle \, p \, \rangle $ of $\mathcal{O}_K$ where $K = \mathbb{Q}(\sqrt{D})$ can be uniquely factored as
\begin{equation}
\langle \, p \, \rangle = \prod_{k=1} \mathfrak{p}_k^{\alpha_k} 
\end{equation}
where $ \mathfrak{p}_k $'s are distinct prime ideals.
Particularly, if $p$ is a rational prime greater than 2, then \cite{marcus1977number}
\begin{equation}\label{(p)}
\langle \, p \, \rangle=p\mathcal{O}_K=\left\{
                     \begin{array}{ll}
                      \mathfrak{p}_1\mathfrak{p}_2, & \hbox{if $\left(\frac{D}{p}\right)=1$;} \\
                      \mathfrak{p}, & \hbox{if $\left(\frac{D}{p}\right)=-1$;} \\
                      \mathfrak{p}^2, & \hbox{if $p | D$.}
                     \end{array}
                  \right.
\end{equation}
where $\mathfrak{p}_1$ and $\mathfrak{p}_2$ are distinct prime ideals, and $\left(\frac{a}{p}\right)$ is the Legendre symbol defined by
\begin{equation*}
\left(\frac{a}{p}\right) = \left\{
\begin{array}{rl}
1 & \text{if for some } x \in \mathbb{Z} \colon\;a\equiv x^2\pmod{p},\\
-1 & \text{otherwise}.
\end{array}
\right.
\end{equation*}

In the first case where $p$ \emph{splits}, $\mathfrak{p}_1$ and $\mathfrak{p}_2$ are distinct prime ideals of norm $p$, thus they are coprime by nature. Particularly in PIDs, the primality of these two ideals can be tested using the criteria given in Section \ref{Prime Elements in Quadratic Fields}.   
Based on $\mathfrak{p}_1$ and $\mathfrak{p}_2$, two coprime ideal lattices can be constructed whereby two subarrays can be allocated on respectively.

\subsection{Coprime Ideal lattices and their matrix representations}

In general, an \emph{ideal lattice} $\Lambda_1=\sigma(\mathcal{I})$ is constructed by canonical embedding from an ideal $\mathcal{I} \subseteq \mathcal{O}_K$, which is a sublattice of the algebraic lattice $\Lambda$ constructed from $\mathcal{O}_K$. 
For example, $7A_2$ is an ideal lattice constructed from the ideal $\langle \, 7 \, \rangle   \, \subset \mathbb{Z}[\omega]$, thus it is a sublattice of $A_2$ constructed from $\mathbb{Z}[\omega]$.

In $\mathbb{Z}[q]$, an integral basis of the principal ideal $\langle \, m \, \rangle $ generated by the quadratic integer $m=m_1+m_2q$ can be calculated as
\begin{equation*}
\begin{aligned}
     m\{ 1 ,  q \} & =\{m_1+m_2q, (m_1+m_2q)q\} \\
                   & =\{m_1+m_2q, -Cm_2+q(m_1-Bm_2)\}.
\end{aligned}
\end{equation*} 
The canonical embedding of a principal ideal maps the elements in the ideal to lattice points, which are similar to that of $ \mathcal{O}_K $ defined in (\ref{Greal}) and (\ref{Gimaginary}) for $D>0$ and $D<0$ respectively.  
Therefore, an ideal lattice denoted as $\sigma(\mathfrak{p})$ that is generated by a principal ideal $\mathfrak{p} = \langle \, m \, \rangle  $ has a generator matrix given by 
\begin{equation}\label{Gm}
  \mathbf{G}_m = \mathbf{G}\mathbf{B}_m, \: \, \text{where}  
\end{equation}
\begin{equation}\label{Bm}
\begin{aligned}
    \mathbf{B}_m = 
    \left(
    \begin{array}{cc}
      m_1 &  -Cm_2\\
      m_2 &  m_1-Bm_2\\
    \end{array}
  \right).
  \end{aligned}
\end{equation}
Here $\mathbf{G}$ is the generator matrix of $\mathcal{O}_K$ that expressed in (\ref{Greal}) or (\ref{Gimaginary}) for real or imaginary quadratic field respectively. $\mathbf{B}_m$ is called the \emph{matrix representation} of $m \in \mathbb{Z}[q]$ and $|\det(\mathbf{B}_m)|=\text{N}(m)$ \cite[Theorem 5.11]{oggier2004algebraic}. Note that $\mathbf{B}_m$ is always an integer matrix by definition, i.e., all entries in $\mathbf{B}_m$ are rational integers. The following two lemmas discuss the properties of $\mathbf{B}_m$ from the perspectives of eigenvectors/eigenvalues and commutativity respectively.  

\begin{lem}\label{eigenvectors}
$(1, \; q)$ and $(1, \; \hat{q})$ are left row eigenvectors of $\mathbf{B}_m$ with eigenvalues $m$ and $\hat{m}$ respectively. 
\end{lem}

\emph{Proof:} 
The row vector $(1, \; q)$ is a left eigenvector of $\mathbf{B}_m$ if $(1, \; q)\mathbf{B}_m=m(1, \; q)$ \cite[Definition 4.2]{hager1988applied}. Substituting (\ref{Bm}) to the left hand side of the equation results in
\begin{eqnarray*}
  & (1, \; q)
   \left(
    \begin{array}{cc}
      m_1 &  -Cm_2\\
      m_2 &  m_1-Bm_2\\
    \end{array}
  \right)  \\
  &  = \big(m_1+m_2q, \; m_1q + m_2(-Bq-C)\big) 
\end{eqnarray*}
As $q$ is the root of $f(X)=0$ where $f(X)$ is given in (\ref{fx2}), it satisfies $q^2+Bq+C=0$. Then the second element in the row vector becomes $m_1q + m_2q^2=(m_1+m_2q)q=mq$. Likewise, substituting $\hat{q}^2=-B\hat{q}-C$ and $\hat{m} = m_1 + m_2\hat{q}$ to $(1, \; \hat{q})\mathbf{B}_m$ yields $(1, \; \hat{q})\mathbf{B}_m= \hat{m} (1, \; \hat{q})$.  
\hfill$\blacksquare$

\begin{lem}\label{commutative matrices}
Any two matrix representations of quadratic integers are commutative.
\end{lem}

\emph{Proof:} 
By the eigendecomposition discussed in Lemma \ref{eigenvectors}, any two matrix representations $\mathbf{B}_m$ and $\mathbf{B}_n$ of two quadratic integers $m$ and $n$ respectively can be factorized as
\begin{equation}\label{decomposition Bm}
    \mathbf{B}_m =\mathbf{Q}^{-1}\mathbf{P}_m\mathbf{Q} \; \text{ and } \; \mathbf{B}_n =\mathbf{Q}^{-1}\mathbf{P}_n\mathbf{Q}
\end{equation}
where 
\begin{equation}\label{Pm}
  \mathbf{Q}=\begin{pmatrix}
    1 & q \\  1 & \hat{q}
  \end{pmatrix},
    \mathbf{P}_m=\begin{pmatrix}
    m & 0  \\  0 & \hat{m}
  \end{pmatrix}
  \: \text{and} \:\,
  \mathbf{P}_n=\begin{pmatrix}
    n & 0  \\  0 & \hat{n}
  \end{pmatrix}
\end{equation}
Therefore, by Lemma \ref{eigenvectors} we can write  
\begin{equation*}
    \begin{aligned}
        & \mathbf{B}_m\mathbf{B}_n= \mathbf{Q}^{-1}\mathbf{P}_m\mathbf{Q}\mathbf{Q}^{-1}\mathbf{P}_n\mathbf{Q}
        = \mathbf{Q}^{-1}\mathbf{P}_m\mathbf{P}_n\mathbf{Q} \\
        & = \mathbf{Q}^{-1}\mathbf{P}_n\mathbf{P}_m\mathbf{Q} = (\mathbf{Q}^{-1}\mathbf{P}_n \mathbf{Q})(\mathbf{Q}^{-1} \mathbf{P}_m\mathbf{Q}) =\mathbf{B}_n\mathbf{B}_m.
    \end{aligned}
\end{equation*}
\hfill$\blacksquare$

In other words, the commutativity of algebraic integers implies the commutativity of the corresponding matrices.
Next, we will relate the coprimality of quadratic integers with the coprimality of their corresponding matrix representations and provide an alternative way of generating coprime lattices from algebraic conjugate pairs.

\subsection{Connection with coprime algebraic integers}

In general, the left coprimality of integers matrices is defined by Bezout's identity \cite{pal2011coprimality,vaidyanathan2011general,Introductiontonumbertheory}:
\begin{deft}\label{ref coprime matrices}
Two integer matrices $\mathbf{B}_m$ and $\mathbf{B}_n$ are left coprime if and only if there exist integer matrices $\mathbf{C}$ and $\mathbf{D}$ such that 
\begin{equation}\label{coprime matrices def}
    \mathbf{B}_m\mathbf{C}+\mathbf{B}_n\mathbf{D} = \mathbf{I}.
\end{equation}
\end{deft}
Likewise, $\mathbf{B}_m$ and $\mathbf{B}_n$ are right coprime if and only if there exist $\mathbf{C}'$ and $\mathbf{D}'$ such that $ \mathbf{C}'\mathbf{B}_m +  \mathbf{D}'\mathbf{B}_n = \mathbf{I}  $.

\begin{ther}\label{integer matrix}
In PIDs, two algebraic integers are coprime if and only if their corresponding matrix representations obtained from canonical embeddings are left coprime.
\end{ther}

\emph{Proof:}
See Appendix \ref{coprime integer and matrix}
\hfill$\blacksquare$

\begin{Corollary}\label{left right coprime}
Two integers matrices generated from embeddings are right coprime if and only if they are left coprime. 
\end{Corollary}
\emph{Proof:} 
See Appendix \ref{appendix left right coprime}
\hfill$\blacksquare$

From Theorem \ref{integer matrix} and Proposition \ref{left right coprime}, the coprimality of two quadratic integers indicates the right and left coprimality of their corresponding matrix representations obtained from embeddings and vice versa. 
Henceforth, we say two lattices are \emph{coprime lattices} if their matrix representations are (left and right) coprime.   
Exploiting this theorem, we will provide conditions on the coprimality of adjugate matrix pairs next.

\subsection{Adjugate Matrix Pairs}

Since the notion of greatest common divisor (GCD) can be generalized to an arbitrary commutative ring, it can be defined in the rings of integers of quadratic fields as well \cite{adhikari2003groups,conrad2008gaussian}. Herein, the concept of GCD is generalized to quadratic integers in PIDs, i.e.,
if $d=\text{GCD}(m,n)$ is a GCD of $m$ and $n$, then all the common divisors of $m$ and $n$ divide $d$ \cite[Definition 6.1.3]{adhikari2003groups}. 
Similarly, the Bezout's identity can also be generalized as if two integers $m , n \in \mathbb{Z}[q]$ are not both equal to 0 then there exist $\alpha , \beta \in \mathbb{Z}[q]$ such that 
\begin{equation*}
    \text{GCD}(m,n) = \alpha m + \beta n.
\end{equation*} 
Given $\text{GCD}(m,n)=u$ where $u$ is the unit in $\mathbb{Z}[q]$ and $\text{N}(u)=1$, $m$ and $n$ are defined as \emph{coprime quadratic integers} \cite[Definition 6.1.4]{adhikari2003groups}. In other words, two quadratic integers $m$ and $n$ are coprime if and only if there exist $\alpha'$ and $\beta'$ such that $m\alpha' + n\beta'=u$. Because $u$ is a unit, $u^{-1}$ always exists. Then Bezout's identity becomes $m\alpha  + n\beta =1$ where $\alpha = \alpha' u^{-1}$ and $\beta = \beta' u^{-1}$. 
Recall that the following facts of GCD hold for $a, b , c \in \mathbb{Z}[q]$:
\begin{enumerate}
  \item $\text{GCD}(a, b) = \text{GCD}(a+\alpha b, b)$, $\forall \alpha \in \mathbb{Z}[q]$;
  \item $\text{GCD}(a, b) = 1$ if and only if $\text{GCD}(a^\alpha, b^\beta)=1,$ $\forall \alpha, \beta \in \mathbb{Z}^+$.
  \item $\text{GCD}(a, bc) = 1$ if and only if $\text{GCD}(a, b) = 1$ and $\text{GCD}(a, c) = 1$. 
\end{enumerate}
These facts can be proved straightforward and will be employed in the following proofs of coprimality.

As mentioned in Section \ref{Quadratic Field and Its Ring of Integers}, the algebraic conjugate of $m$ denoted by $\hat{m}$ is also in $\mathbb{Z}[q]$ and can be written as $m_1+m_2\hat{q}$. The matrix generated by $\hat{m}$ is the same as the adjugate of $\mathbf{B}_m$, which is the transpose of the cofactor matrix of $\mathbf{B}_m$, i.e., $adj(\mathbf{B}_m)_{kj}=(-1)^{k+j}\mathbf{M}_{jk}$ where 
$\mathbf{M}_{jk}$ is the determinant of the matrix that results from deleting row $k$ and column $j$ of $\mathbf{B}_m$. Therefore, in the dimension of two, the adjugate of $\mathbf{B}_m$ is 
\begin{equation}\label{Gmhat}
    \mathbf{B}_{\hat{m}} = \left(
    \begin{array}{cc}
      m_1-Bm_2 &  Cm_2\\
      -m_2 &  m_1\\
    \end{array}
  \right).
\end{equation}

\begin{ther}\label{adjugate matrices}
Using the notations above, two adjugate 2-by-2 matrices $\mathbf{B}_m$ and $\mathbf{B}_{\hat{m}}$ are coprime if and only if 
\begin{enumerate}[label=(\alph*)]
\item $\text{GCD}(m_1,m_2)=1$ and $\text{GCD}(2m_1+m_2, \, 4C-1)=1$, for $B =  - 1$,
\item $\text{GCD}(m_1,m_2)=1$ and $\text{GCD}(m_1, \, C)=1$, for $B =  0$ and $C$ is even,
\item $\text{GCD}(m_1+m_2,m_1-m_2)=1$ and $\text{GCD}(m_1, \, C)=1$ for $B =  0$ and $C$ is odd.
\end{enumerate}
 \end{ther}
 
\emph{Proof}: 
See Appendix \ref{coprime conditions}
\hfill$\blacksquare$

It is worth to notice that according to Theorem 1, Theorem 2 also provides the coprime conditions of two algebraic conjugates $m = m_1 + m_2q$ and $\hat{m} = m_1 + m_2\hat{q}$ where $q$ and $\hat{q}$ are given in (\ref{q}) and (\ref{hatq}) respectively.   
For illustration purposes, two examples of algebraic conjugate integers are given, providing two classes of coprime matrices. 

\begin{Corollary}\label{Gaussian conjugate}
A Gaussian integer $m$ and its conjugate are relatively prime if and only if $\text{GCD}(m_1+m_2,m_1-m_2)=1$. 
\end{Corollary}

\emph{Proof}:
The minimum polynomial of the ring of Gaussian integers is $X^2+1=0$ with the basis $\{ 1, i\}$. By Theorem \ref{adjugate matrices}, $B=0$ and $C=1$ match the assumptions of case (c), thus
the coprimality condition becomes $\text{GCD}(m_1+m_2,m_1-m_2)=1$ and $\text{GCD}(m_1,1)=1$. Note that $\text{GCD}(m_1,1)=1$ holds for all $m_1 \in \mathbb{Z}$. 
\hfill$\blacksquare$

\begin{Corollary}\label{E conjugate}
An Eisenstein integer and its conjugate are relatively prime if and only if $\text{GCD}(m_1,m_2)=1$ and $\text{GCD}(m_1-m_2,3)=1$.  
\end{Corollary}

\emph{Proof}:
Likewise, in the case of $\mathbb{Z}[\omega]$, the coefficients become $C=1$ and $B=-1$, which can be addressed to case (a) in Theorem \ref{adjugate matrices}. The coprime conditions are $\text{GCD}(m_1,m_2)=1$ and $\text{GCD}(2m_1+m_2,3)=1$. By Fact (1), $\text{GCD}(m_1,m_2)=1$ is equivalent to $\text{GCD}(2m_1+m_2,m_1)=1$, which can be combined with the second condition by Fact (3), i.e., $\text{GCD}(2m_1+m_2,3m_1)=1$. Applying Fact (1) again results in $\text{GCD}(3m_1-(2m_1+m_2),3m_1)=\text{GCD}(m_1-m_2,3)=1$, which holds if and only if both $\text{GCD}(m_1,m_2)=1$ and $\text{GCD}(m_1-m_2,3)=1$ hold.       
\hfill$\blacksquare$

\emph{Remark:} By Theorem \ref{adjugate matrices}, the class of coprime matrix pairs is enriched to any matrices obtained from quadratic integers.
By exploiting the bijective mappings between algebraic integers and integer matrices, \cite[Theorem 2]{vaidyanathan2011general} can be viewed as the coprimality of two algebraic conjugates and proved by the generalized GCD, i.e.,
\begin{equation*}
    \begin{aligned}
       &  \text{GCD}(m_1 + m_2 q, \, m_1 + m_2 \hat{q} )    \\
       & = \text{GCD}\big((m_1 + m_2 q)(m_1 + m_2 \hat{q} ), \, 2m_1 + m_2 (q+ \hat{q})\big) \\
       & = \text{GCD}(\text{N}(m), 2m_1 - Bm_2).
    \end{aligned}
\end{equation*}
According to the coprimality relation between quadratic integers and matrices stated in Theorem \ref{integer matrix}, some useful classes of coprime matrices such as skew-circulant adjugates derived in \cite{pal2011coprimality} can be viewed as a case of Corollary \ref{Gaussian conjugate} with canonical embedding, i.e., the following two integer matrices
\begin{equation*}
  \mathbf{B}_m=\begin{pmatrix}
    m_1 & -m_2\\ m_2  & m_1 
  \end{pmatrix}
  \quad \text{and} \quad
  \mathbf{B}_{\hat{m}}=\begin{pmatrix}
     m_1 & m_2\\ -m_2  & m_1 
  \end{pmatrix}
\end{equation*}
are coprime if and only if $\text{GCD}(m_1+m_2, \, m_1 - m_2)=1$. Similarly, an alternative way of describing Corollary \ref{E conjugate} would be as follows:

Two integer matrices
\begin{eqnarray*}
  & \mathbf{B}_m=\begin{pmatrix}
    m_1 & -m_2\\ m_2  & m_1+m_2 
  \end{pmatrix}
  \quad \text{and} \\
  &
  \mathbf{B}_{\hat{m}}=\begin{pmatrix}
     m_1 + m_2 & m_2\\ -m_2  & m_1 
  \end{pmatrix}
\end{eqnarray*}
are coprime if and only if $\text{GCD}(m_1, \, m_2)=1$ and $\text{GCD}(m_1 - m_2, \, 3)=1$, i.e., $m_1$ and $m_2$ are relatively prime with their difference being not divisible by 3.   

\section{Design of CRT-based Sparse Arrays}\label{Design of CRT-based Sparse Arrays}

We have proved that the coprimality of quadratic integers in PIDs is a necessary and sufficient condition of the coprimality of the corresponding matrices obtained from canonical embeddings.  
In this section, we will briefly review the generalized CRT based on \cite[Appendix 1]{marcus1977number}, and then propose a design method for coprime arrays based on CRT over rings of quadratic integers where the sensors are deployed by the use of coprime lattices generated by coprime ideals. 
For illustrative purposes, CRT is employed over the ring of Gaussian integers and the ring of Eisenstein integers respectively as examples.    

To begin with, we extend the modulo operation to ideals, and define the sum and the product of ideals as follows:
\begin{deft}
Let $\mathcal{I}$ be an ideal of a ring $R$. Given $x,y \in R$, $x$ is congruent to $y$ modulo $\mathcal{I}$, i.e., $x \equiv y \mod{\mathcal{I}}$ if and only if 
\begin{equation}
    x - y \in \mathcal{I}.
\end{equation}
\end{deft}

\begin{deft}
Let $\mathcal{I}$ and $\mathcal{J}$ be two ideals of the ring $R$. The sum of $\mathcal{I}$ and $\mathcal{J}$ is the ideal
\[
\mathcal{I} + \mathcal{J} = \{x+y, x\in \mathcal{I}, y \in \mathcal{J}\},
\]
and their product is the ideal
\[
 \mathcal{I} \mathcal{J} = \{\sum{x_k y_j}, x_k \in \mathcal{I}, y_j \in \mathcal{J}\}.
\]
\end{deft}

The quotient ring is defined as the set that contains all the cosets of the  ideal $\mathcal{I}$, i.e., $R/\mathcal{I} = \{r + \mathcal{I}, r \in R\}$. 
Let ideals $\mathcal{I}$ and $\mathcal{J}$ be relatively prime in a commutative ring $R$, then  $\mathcal{I} + \mathcal{J} = R$. 
For example, given $\mathcal{I}= \langle \, 3 \, \rangle  $ and $ \mathcal{J} = \langle \, 5 \, \rangle $ as two coprime ideals of $\mathbb{Z}$, $\mathcal{I}+\mathcal{J} = \langle \, 3 \, \rangle + \langle \, 5 \, \rangle = \langle \, 3\cdot2 + 5\cdot(-1) \, \rangle = \langle \, 1 \, \rangle  = \mathbb{Z} $, $\mathcal{I}\mathcal{J}=\langle \, 15 \, \rangle $, and $R/\mathcal{I} = \mathbb{Z}/3\mathbb{Z} = \{ 0, 1, 2 \}$ which is an equivalence class with $[x] = [y]$ if and only if $x-y \in  \langle \, 3 \, \rangle $.

The Chinese Remaindering Theorem \cite{marcus1977number} asserts that there is a ring isomorphism
\begin{equation}\label{ring isomorphism}
    R/\mathcal{IJ} \simeq R/\mathcal{I} \times R/\mathcal{J}.
\end{equation}
This implies that for all $ a_k \in R/\mathcal{I}$ and $b_j \in R/\mathcal{J}$ there exists $z \in R/\mathcal{IJ}$ such that
\begin{equation}\label{zaibi}
\begin{aligned}
& z  \quad \equiv  \quad a_k \pmod{\mathcal{I}} \: \, \text{and} \\
& z  \quad \equiv \quad  b_j \pmod{\mathcal{J}},
\end{aligned}
\end{equation}
which can also be proved as follows: from coprimality that $\mathcal{I} + \mathcal{J} = R$, there exist $x_k\in \mathcal{I}$ and $y_j \in \mathcal{J}$ such that $x_k+ y_j =1$.
For all $ a_k \in R/\mathcal{I}$ and $b_j \in R/\mathcal{J}$, it can be readily verified that every pair $(a_k, b_j)$ forms the solution
\begin{equation}
z \equiv x_k b_j + y_j a_k \pmod{\mathcal{IJ}}.
\end{equation}
We may check that
\begin{eqnarray*}
 z &\equiv& y_j a_k \equiv x_k a_k + y_j a_k = (x_k + y_j) a_k \equiv a_k \pmod{\mathcal{I}} \\
 z &\equiv& x_k b_j \equiv x_k b_j + y_j b_j = (x_k + y_j) b_j \equiv b_j \pmod{\mathcal{J}}.
\end{eqnarray*}
The pair $(x_k, y_j)$ serves as a ``CRT basis" which can be chosen as the basis of prime ideals in $\mathbb{Z}[q]$.
With this basis, the mapping from $R/\mathcal{I} \otimes R/\mathcal{J}$ to $R/\mathcal{IJ}$ is bijective, i.e., all solutions of $z$ are identical given the different pairs $(a_k,b_j)$, which leads to the definition of CRT arrays and its cross-difference and sum coarrays:

\begin{deft}[CRT arrays] \label{crt arrays}
Given two coprime ideals $\mathcal{I}$ and $\mathcal{J}$ in ring $R$, a CRT-based array is defined as:
\begin{equation}\mathcal{Z} =  \sigma (\mathcal{I}) /  \sigma (\mathcal{I}\mathcal{J}) \cup  \sigma (\mathcal{J}) /  \sigma (\mathcal{I}\mathcal{J}),   
\end{equation}
where $\sigma(\mathcal{I})$ denotes the canonical embedding of $\mathcal{I}$ and same with $\mathcal{J}$. 
\end{deft}
\begin{deft} [Cross-difference coarrays of CRT arrays]
The cross-difference coarray $\mathcal{D}$ generated by an CRT array is given by:
\begin{equation*}
\mathcal{D} = \{ \mathbf{z}_1 - \mathbf{z}_2 \mid  \mathbf{z}_1 \in \sigma (\mathcal{I}) / \sigma(\mathcal{I}\mathcal{J}),  \mathbf{z}_2 \in \sigma(\mathcal{J}) / \sigma(\mathcal{I}\mathcal{J}) \}. 
\end{equation*}
\end{deft}
\begin{deft} [Sum coarray of CRT arrays]
The sum coarray $\mathcal{S}$ generated by an CRT array can be expressed as:
\begin{equation*}
\mathcal{S} = \{ \mathbf{z}_1 + \mathbf{z}_2 \mid  \mathbf{z}_1 \in \sigma (\mathcal{I}) / \sigma(\mathcal{I}\mathcal{J}),  \mathbf{z}_2 \in \sigma(\mathcal{J}) / \sigma(\mathcal{I}\mathcal{J}) \}. 
\end{equation*}
\end{deft}
Note that because of the symmetry of the ideal lattices, $\mathcal{D}$ is identical to $\mathcal{S}$.
From the point of view that regards lattices as sets of points, $\sigma (\mathcal{I}) / \sigma(\mathcal{I}\mathcal{J})$ corresponds to $\mathcal{I} / (\mathcal{I}\mathcal{J})$ in number fields.    

According to \cite{marcus1977number}, the ring isomorphism (\ref{ring isomorphism}) holds over any commutative ring, therefore over all PIDs where the ideals can be obtained from the prime decomposition (\ref{(p)}). 
Given $\mathcal{I}=\mathfrak{p}_1= \langle \, m \, \rangle  $ and $\mathcal{J}=\mathfrak{p}_2= \langle \, n \, \rangle$ where $m,n \in \mathbb{Z}[q]$, the canonical embedding $ \sigma(\mathfrak{p}_1) $ of $ \mathfrak{p}_1 $ is given in (\ref{Gm}) and similar with $ \sigma(\mathfrak{p}_2) $.  
The product of these two ideals forms a principal ideal as well, i.e., $\mathcal{I}\mathcal{J}= \langle \, mn \, \rangle  = \langle \, p \, \rangle $.
With the notations above, expressions for the number of sensors and the achievable DOF can be derived as follows: 
\begin{Proposition}\label{CRT array}
If $\mathcal{I}$ and $\mathcal{J}$ that are used to allocate sensors are decomposed from $\langle \, p \, \rangle$, the total number of physical sensors is $2p-1$ and its maximum DOF is $p^2$.   
\end{Proposition}

\emph{Proof}: 
By assumption $\mathcal{I} \mathcal{J}= \langle \, p \, \rangle  = pR$, the number of elements in $\mathcal{I}/p R $ is $p$ which is the same as $\mathcal{J}/p R$ \cite[Definition 3.12]{oggier2004algebraic}.
Since the only identical element that they share is $\mathbf{0}$, the total number of nonidentical elements in $ (\mathcal{I} / p R) \cup (\mathcal{J}/ p R ) $ is $2p-1$. 

Define the maximum DOF as the maximum number of degrees of freedom that the array can achieve, i.e., the total number of identical elements in the coarray. 
According to the ring isomorphism of the generalized CRT given in (\ref{ring isomorphism}), all the difference/sum vectors generated by coprime lattices are nonidentical, thus the total number of elements in $(\mathcal{I} + \mathcal{J})/pR$ can be written as:
\begin{equation}
 | \mathcal{I} /p R | \cdot |  \mathcal{J}  /p R |=p^2,
\end{equation}
where $|\cdot|$ is the cardinality of a set. Note that as canonical embedding $\sigma(\cdot)$ is bijective, the number of lattice points in $\sigma(\mathcal{J}) / \sigma(\mathcal{I}\mathcal{J})$ is the same as the number of elements in $\mathcal{J} / \mathcal{I}\mathcal{J}$
\hfill$\blacksquare$

The rest of this section will demonstrate the proposed design method by Chinese Remaindering over PIDs in quadratic fields, namely over $\mathbb{Z}[i]$ and $\mathbb{Z}[\omega]$.

\subsection{Chinese Remaindering over $\mathbb{Z}[i]$}

In the ring of Gaussian integers, we look for the $p$ such that $D=-1$ is a quadratic residue:
\[
x^2  \equiv -1  \pmod{p}
\]
for some $x \in \mathbb{Z}$. The first few solutions are $2^2  \equiv -1  \pmod{5}$, $5^2  \equiv -1  \pmod{13}$, and so forth. By performing the prime decomposition (\ref{(p)}), these rational primes can be decomposed into prime ideals as
$  \langle \, 5 \, \rangle  =  \langle \, 2+i \, \rangle  \langle \, 2-i \, \rangle $, and $ \langle \, 13 \, \rangle  =  \langle \, 3+2i \, \rangle    \langle \, 3-2i \, \rangle $.
Here all the quadratic integers are Gaussian primes as stated in the criteria given in Section \ref{Prime Elements in Quadratic Fields}.
Alternatively, it can be checked that all pairs are relatively prime according to Corollary \ref{Gaussian conjugate}.
Let us take the example of $p=5$ to demonstrate the design procedure. 
$p=5$ yields two prime ideals $\mathfrak{p}_1=\langle \, 2+i \, \rangle $ and $\mathfrak{p}_2= \langle \, 2-i \, \rangle $, whose corresponding matrices are coprime as well by Theorem \ref{integer matrix}. 
As $\{1,i\}$ is the integral basis of $\mathbb{Z}[i]$, an integral basis of $\langle \, 2+i \, \rangle $ can be calculated as
\[
(2+i)\{1,i\}=\{2+i,-1+2i\}.
\]
Since the minimum polynomial over $\mathbb{Z}[i]$ is $X^2+1$ ($B=0$ and $C=1$), by canonical embedding given as (\ref{Gm}) and (\ref{Bm}), the generator matrix of $\langle \, 2+i \, \rangle$ is
\begin{eqnarray}\label{G1-Gaussian}
  \mathbf{G}_{(2+i)} = \left(
    \begin{array}{cc}
      2 &  -1\\
      1 & 2 \\
    \end{array}
  \right).
\end{eqnarray}
Notice that because $\mathbf{G}=\mathbf{I}$, the generator matrix of $\langle \, 2+i \, \rangle$ is identical to its matrix representation, i.e., $\mathbf{G}_{(2+i)}=\mathbf{B}_{(2+i)}$. Analogously, an integral basis of $\langle \, 2-i \, \rangle$ is given by
\[
(2-i)\{1,i\}=\{2-i,1+2i\},
\]
whose generator matrix is
\begin{eqnarray}\label{G2-Gaussian}
  \mathbf{G}_{(2-i)} = \left(
    \begin{array}{cc}
      2 &  1\\
      -1 & 2 \\
    \end{array}
  \right).
\end{eqnarray}
The determinant of $\mathbf{G}_{(2+i)}$ is equivalent to the norm of $\langle \, 2+i \, \rangle$ and same with $\mathbf{G}_{(2-i)}$ and $\langle \, 2-i \, \rangle$ \cite{oggier2004algebraic}.
By the definition of the norm (\ref{normm}), it can be proved straightforward that if $ \langle \, p \, \rangle =\langle \, 2+i \, \rangle \langle \, 2-i \, \rangle $, $ \text{N}(\langle \, 2+i \, \rangle)\text{N}(\langle \, 2-i \, \rangle) = \text{N}(p) $ and since $p$ is a prime number, $ \text{N}(p) = p^2 $ has and only has three divisors, namely $1$, $p$, and $p^2$. According to the decomposition shown in (\ref{(p)}), both $\langle \, 2+i \, \rangle$ and $\langle \, 2-i \, \rangle$ are not units.    
This implies $ | \det(\mathbf{G}_{(2+i)}) | = | \det(\mathbf{G}_{(2-i)}) | =  \text{N}(\langle \, 2+i \, \rangle)  = \text{N}(\langle \, 2-i \, \rangle) = p = 5$.

 \begin{figure}[tb]
  \begin{center}
      \subfigure[]{\includegraphics[height=2.2in]{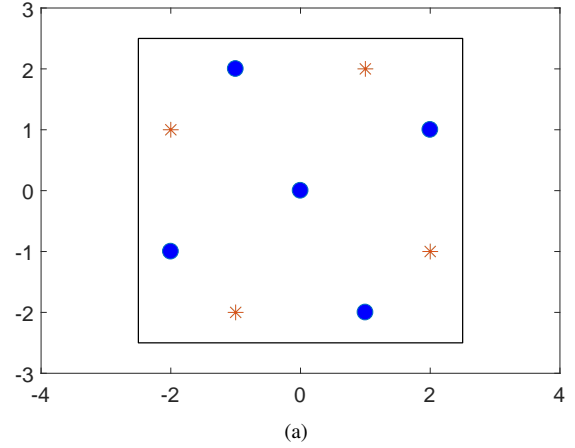}\label{g5holepositions}}
      \subfigure[]{\includegraphics[height=2.2in]{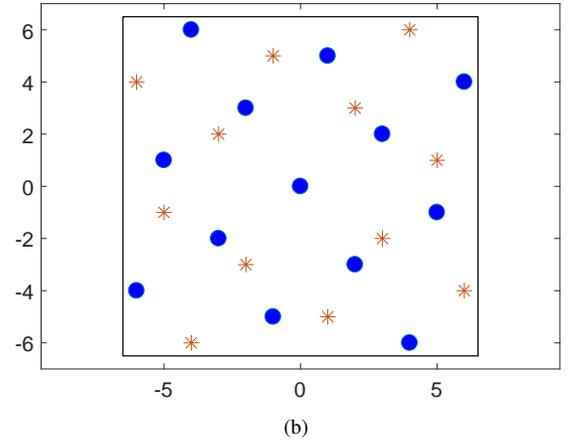}\label{g13holepositions}}
    \caption{$\mathbb{Z}^2$ arrays in the Voronoi cells (black polygons) constructed from decomposition over Gaussian integers of $p=5$ (a) and $p=13$ (b) with the first subarray ($\mathbf{G}_{(2+i)}\mathbf{g_1}$) in red stars and the second subarray ($\mathbf{G}_{(2-i)}\mathbf{g_2}$) in blue dots.}
    \label{sensorposition}
  \end{center}
\vspace{-1em}
\end{figure}
 
Using the matrix representation, the cross-difference coarray consisting of vectors can be defined by 
\begin{equation*}
\mathcal{D}_G = \{ \mathbf{d}_g: \mathbf{d}_g = \mathbf{G}_{(2-i)} \mathbf{g}_1 - \mathbf{G}_{(2+i)} \mathbf{g}_2 \},
\end{equation*}
where $\mathbf{g}_1 \in \mathbb{Z}^2/\sigma\big(\langle \, 2+i \, \rangle\big)$ and $\mathbf{g}_2 \in \mathbb{Z}^2/\sigma\big(\langle \, 2-i \, \rangle\big)$. $\mathbf{G}_{(2+i)}$ and $\mathbf{G}_{(2-i)}$ are given by (\ref{G1-Gaussian}) and (\ref{G2-Gaussian}) respectively and they are coprime because $ 2+ i $ and $ 2 - i $ are coprime integers (Theorem \ref{integer matrix}).  
According to the ring isomorphism of the generalized CRT, with $R = \mathbb{Z}$, $\mathcal{I}= \langle \, 2+i \, \rangle $ and $ \mathcal{J} = \langle \, 2-i \, \rangle$, it can be readily calculated that $\mathcal{IJ}=\langle \, 5 \, \rangle = 5\mathbb{Z}$ and thus $\mathbf{d}_g \in \mathbb{Z}^2/5\mathbb{Z}^2$, yielding an array of $5^2=25$ degrees of freedom according to Proposition \ref{CRT array}.
The locations of the elements of the first subarray are given by
\[
\mathbf{G}_{(2-i)} \mathbf{g}_1 \in \sigma \big(\langle \, 2-i \, \rangle \big)/5\mathbb{Z}^2,
\]
while those of the second subarray are given by
\[
\mathbf{G}_{(2+i)} \mathbf{g}_2 \in \sigma\big(\langle \, 2+i \, \rangle\big)/5\mathbb{Z}^2.
\]
Therefore, $\mathcal{Z}=\sigma\big(\langle \, 2-i \, \rangle\big)/5\mathbb{Z}^2 \cup  \, \sigma\big(\langle \, 2+i \, \rangle\big)/5\mathbb{Z}^2$ and only 9 elements are actually used in the sparse array by Proposition \ref{CRT array}.

Another example that comprises more sensors is $p=13$, where two coprime ideals $ \langle \, 3+2i \, \rangle $ and $\langle \, 3-2i \, \rangle $ can be obtained from (\ref{(p)}). The generator matrices of the corresponding ideal lattices are given by
\begin{equation*}
  \mathbf{G}_{(3+2i)}=\begin{pmatrix}
    3 & -2\\        2&3
  \end{pmatrix}
  , \: \text{and} \:\,
  \mathbf{G}_{(3-2i)}=\begin{pmatrix}
    3&2\\        -2&3
  \end{pmatrix}.
\end{equation*}
This coprime array produces $13^2=169$ DOFs from $13\times2-1=25$ physical sensors according to Proposition \ref{CRT array}.

Since $\mathbb{Z}[i]$ is isomorphic to polynomial ring $\mathbb{Z}[x]/(x^2+1)$ that gives rise to skew-circulant matrices, the sensor arrays obtained from skew-circulant matrices in \cite{vaidyanathan2011theory} can be viewed as CRT arrays over $\mathbb{Z}[i]$.  
Symmetric Voronoi regions $\mathcal{V}(p\mathbb{Z}^2)$ defined in (\ref{Vorinoi regin}) are used to modulo these sensors corresponding to algebraic integers in $\mathbb{Z}[i]$, as depicted in Fig. \ref{g5holepositions} and Fig. \ref{g13holepositions} for $p=5$ and $p=13$ respectively.

\subsection{Chinese Remaindering over $\mathbb{Z}[\omega]$}\label{CRT over Eisenstein}

 \begin{figure}[tb]
  \begin{center}
    \subfigure[]{\includegraphics[height=2.2in]{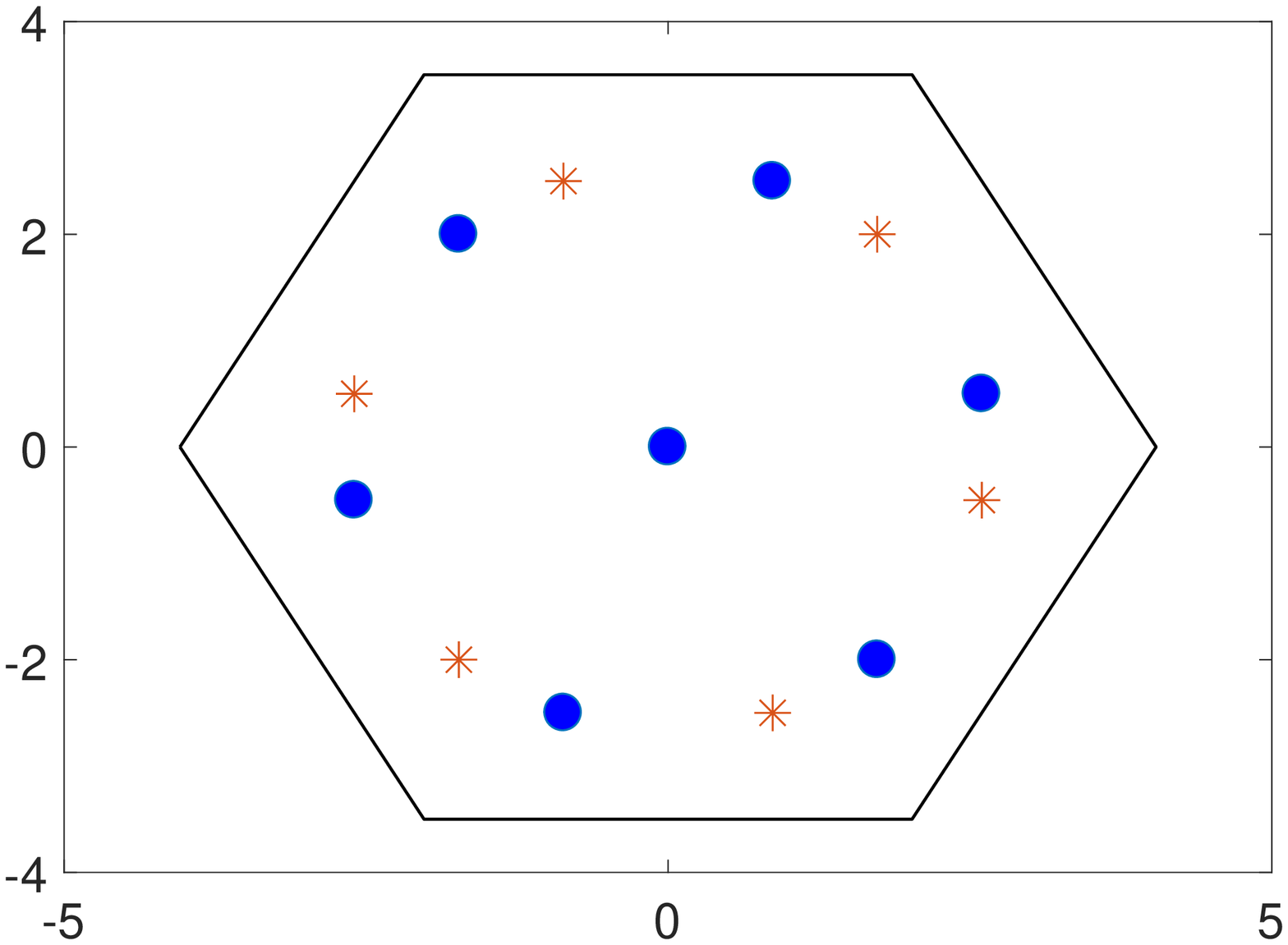}\label{e7holepositions}}
     \subfigure[]{\includegraphics[height=2.2in]{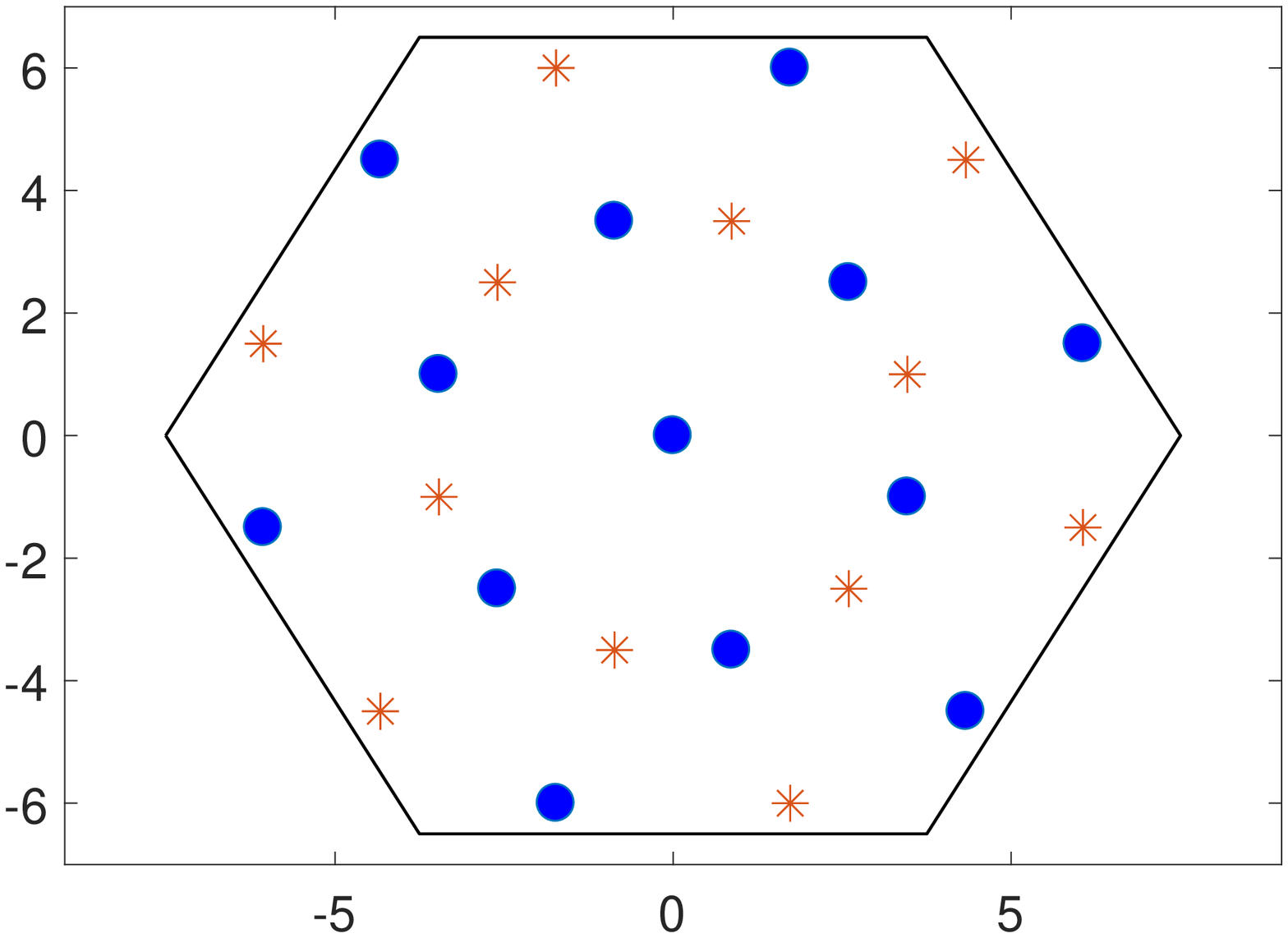}\label{e13holepositions}}
    \caption{$A_2$ arrays in the Voronoi cells (black polygons) constructed from decomposition of $p=7$ (a) and $p=13$ (b) over Eisenstein integers.}
  \end{center}
\vspace{-2em}
\end{figure}

To derive the prime ideals in the ring of Eisenstein integers, we shall aim for $p$ such that $D=-3$ is a quadratic residue:
\[
x^2  \equiv -3  \pmod{p},
\]
for some $x \in \mathbb{Z}$. For example, solutions can be $2^2  \equiv -3  \pmod{7}$, $6^2  \equiv -3  \pmod{13}$, and so forth. By performing ideal decomposition (\ref{(p)}), these rational primes can be decomposed into prime ideals as:
 $\langle \,  7 \, \rangle = \langle \, 2+\sqrt{3}i \, \rangle \langle \, 2-\sqrt{3}i \, \rangle$ and $\langle13 \, \rangle = \langle1+2\sqrt{3}i \, \rangle  \langle \, 1-2\sqrt{3}i \, \rangle$.
With $p=7$, the two prime ideals decomposed from $\langle \, 7 \, \rangle $ are $\mathfrak{p}_1= \langle \, 2+\sqrt{3}i \, \rangle$ and $\mathfrak{p}_2= \langle \, 2-\sqrt{3}i \, \rangle $, where $2+\sqrt{3}i$ and $2-\sqrt{3}i$ are prime elements by the criteria given in Section \ref{Prime Elements in Quadratic Fields}. Because $\mathfrak{p}_1$ and $\mathfrak{p}_2$ are algebraic conjugate of each other ($2+\sqrt{3}i=1+2\omega$ and $2-\sqrt{3}i=1+2\hat{\omega}$), 
it can also be checked that these two conjugate Eisenstein integers are coprime according to Corollary \ref{E conjugate} with $m_1=1$ and $m_2=2$. Similar to Gaussian integers, $ \langle \, 2+\sqrt{3}i \, \rangle$ in $\mathbb{Z}[\omega]$ has an integral basis represented by 
\[
(2+\sqrt{3}i)\{1,\omega\}=\bigg\{ 2+\sqrt{3}i,\frac{-1+3\sqrt{3}i}{2} \bigg\}
\]
whose corresponding generator matrix is
\begin{eqnarray*}\label{G1-Eisenstein}
  \mathbf{G}_{(2+\sqrt{3}i)} &=& \left(
    \begin{array}{cc}
      2 &  -\frac{1}{2}\\
      \sqrt{3} & \frac{3\sqrt{3}}{2} \\
    \end{array}
  \right)  \\
   &=& \left(
    \begin{array}{cc}
      1 &  \frac{1}{2}\\
      0 & \frac{\sqrt{3}}{2} \\
    \end{array}
  \right) \underbrace{\left(
    \begin{array}{cc}
      1 &  -2\\
      2 & 3 \\
    \end{array}
  \right),}_{\mathbf{B}_{(2+\sqrt{3}i)}}
\end{eqnarray*}
and the integral basis of $ \langle \, 2-\sqrt{3}i \, \rangle$ is given by
\[
(2-\sqrt{3}i) \{1,\omega\}=\bigg\{ 2-\sqrt{3}i,\frac{5+\sqrt{3}i}{2} \bigg\}
\]
with the generator matrix:
\begin{eqnarray*}\label{G2-Eisenstein}
  \mathbf{G}_{(2-\sqrt{3}i)} &=& \left(
    \begin{array}{cc}
      1 &  \frac{1}{2}\\
      0 & \frac{\sqrt{3}}{2} \\
    \end{array}
  \right) \underbrace{\left(
    \begin{array}{cc}
      3 &  2\\
      -2 & 1 \\
    \end{array}
  \right).}_{\mathbf{B}_{(2-\sqrt{3}i)}}
\end{eqnarray*}
Here the matrices $\mathbf{B}_{(2+\sqrt{3}i)}$ and $\mathbf{B}_{(2-\sqrt{3}i)}$ are the matrix representations of ideals $ \langle \, 2+\sqrt{3}i \, \rangle$ and $ \langle \, 2-\sqrt{3}i \, \rangle$ respectively and they are coprime according to Theorem \ref{integer matrix}. Similar to the Gaussian case, it can be verified that $|\det(\mathbf{B}_{(2+\sqrt{3}i)})|=|\det(\mathbf{B}_{(2-\sqrt{3}i)})|= \text{N}( \langle \, 2+\sqrt{3}i \, \rangle)=\text{N}( \langle \, 2-\sqrt{3}i \, \rangle)= p=7$. 
With the notations above, the locations of the elements of the first subarray are given by
$
\mathbf{G}_{(2-\sqrt{3}i)} \mathbf{e}_1 \in \sigma \big( \langle \, 2-\sqrt{3}i \, \rangle \big)/7A_2,
$
while those positions of the second subarray are
$
\mathbf{G}_{(2+\sqrt{3}i)} \mathbf{e}_2 \in \sigma \big( \langle \, 2+\sqrt{3}i \, \rangle \big)/7A_2, 
$ 
where $\mathbf{e}_1 \in A_2/ \sigma \big(  \langle \, 2+\sqrt{3}i \, \rangle \big)$, and $\mathbf{e}_2 \in A_2/\sigma\big(  \langle \, 2-\sqrt{3}i \, \rangle \big)$.
The elements in the cross-difference coarray are defined in the same form as the $\mathbb{Z}^2$ array, i.e.,
$
\mathbf{d}_e = \mathbf{G}_{(2-\sqrt{3}i)} \mathbf{e}_1 - \mathbf{G}_{(2+\sqrt{3}i)} \mathbf{e}_2.
$
By substituting $R=A_2$, $\mathcal{I}= \langle \, 2+\sqrt{3}i \, \rangle $ and $ \mathcal{J} = \langle \, 2-\sqrt{3}i \, \rangle $ to (\ref{ring isomorphism}), the generalized CRT guarantees $\mathbf{d}_e \in A_2/7A_2$, yielding an array of $49$ degrees of freedom with $13$ sensors according to Proposition \ref{CRT array}.

\begin{table*}
 \caption{Hole-free symmetric CRT array design}
  \hrule
  \begin{algorithmic}[1]
    \REQUIRE A PID $\mathbb{Q}(\sqrt{D})$. 
     \STEP  
      \STATE Calculate a rational integer $p \in \mathbb{Z}$ such that Legendre symbol $\big( \frac{D}{p} \big) =1$.
      \STATE Compute the integral basis $\{1,q\}$ of $\mathbb{Z}[q]$ as $q = -\frac{1}{2}B + \frac{1}{2}\sqrt{B^2-4C}$ where $B=0$ and $C=-D$ if $D \not\equiv 1 \pmod{4}$, and $B=-1$ and $C=\frac{1-D}{4}$ if $D \equiv 1 \pmod{4}$.
      \STATE Decompose $ \langle \, p \, \rangle  $ into two coprime ideals $\mathfrak{p}_1=  \langle \, m_1 + m_2q \, \rangle $ and $\mathfrak{p}_2=  \langle \,  n_1+n_2q  \, \rangle$ of $\mathbb{Q}(\sqrt{D})$.  
      \STATE Compute generator matrices as
\begin{eqnarray*}
   \mathbf{G}_1 = \mathbf{G} \left(
    \begin{array}{cc}
       m_1 &  -Cm_2\\
      m_2 &  m_1-Bm_2\\
    \end{array}
  \right) \qquad  \text{and} \qquad 
  \mathbf{G}_2 = \mathbf{G} \left(
    \begin{array}{cc}
       n_1 &  -Cn_2\\
      n_2 &  n_1-Bn_2\\
    \end{array}
  \right),
\end{eqnarray*} 
where $\mathbf{G}$ is defined in (\ref{Greal}) for $D>0$ and (\ref{Gimaginary}) for $D<0$.
      \STATE The sensors are allocated on a 2D space where the positions are given by $\mathbf{G}_1 \mathbf{x}_2 $ and $\mathbf{G}_2 \mathbf{x}_1 $ with $\mathbf{x}_1 \in \sigma(\mathbb{Z}[q])/2\sigma(\mathfrak{p}_1) $ and $\mathbf{x}_2 \in \sigma(\mathbb{Z}[q])/\sigma(\mathfrak{p}_2) $.
  \end{algorithmic}
   \hrule
    \label{hole free table}
\end{table*}

Similarly, with a larger array aperture such as $p=13$, the two prime ideals decomposed from $\langle \, 13 \, \rangle $ are $\langle \, 1+2\sqrt{3}i \, \rangle = \langle \,-1+4\omega \, \rangle $ and $ \langle \,  1-2\sqrt{3}i \, \rangle =\langle \, -1 + 4 \hat{\omega}\, \rangle $, which are coprime according to  Corollary \ref{E conjugate} with $m_1=-1$ and $m_2=4$.
The generator matrices of the corresponding ideal lattices are given by
\begin{equation*}
\begin{aligned}
    &  \mathbf{G}_{(1+2\sqrt{3}i)}=\begin{pmatrix}
    1 & -\frac{5}{2}\\   2\sqrt{3}&\frac{3\sqrt{3}}{2}
  \end{pmatrix} 
   \: \text{and} \\
    &  \mathbf{G}_{(1-2\sqrt{3}i)}=\begin{pmatrix}
    1&\frac{7}{2}\\ -2\sqrt{3} &-\frac{\sqrt{3}}{2}
  \end{pmatrix}.
\end{aligned}
\end{equation*}
Here the set of the difference coarray is $ A_2/13A_2$ Thus this array provides $169$ DOF with $25$ sensors. The subarrays are allocated on the following two ideal lattices:
$$
 \sigma\big(\langle \, 1-2\sqrt{3}i \, \rangle\big)/13A_2,
\; \text{and} \; \sigma\big( \langle \, 1+2\sqrt{3}i \, \rangle\big)/13A_2.
$$
Similar to the Gaussian cases, Voronoi regions $\mathcal{V}(pA_2)$ are employed to allocate sensors, yielding more compact arrays. The examples of $A_2$ array configurations are shown in Fig. \ref{e7holepositions} and Fig. \ref{e13holepositions} for $p=7$ and $p=13$ respectively. To highlight the shape of hexagonal Voronoi cell for illustrative purposes, all $A_2$ based arrays are rotated $90$ degrees counterclockwise henceforth.

When all antennas act like receivers, the array configuration given in Definition \ref{crt arrays} can be redefined equivalently as a set that consists of sensor locations given by vectors, i.e., 
\begin{equation*}
 \begin{aligned}
   \mathcal{Z} = &\{  \mathbf{z} = \mathbf{G}_2 \mathbf{x}_1 \; | \;    \mathbf{x}_1 \in \sigma (\mathbb{Z}[q])/\sigma(\mathfrak{p}_1) \} \\
                 &\cup  \{ \mathbf{z} = \mathbf{G}_1 \mathbf{x}_2 \; | \;  \mathbf{x}_2 \in  \sigma(\mathbb{Z}[q])/\sigma(\mathfrak{p}_2) \},
 \end{aligned}   
\end{equation*}
where $\mathbf{G}_1$ and $\mathbf{G}_2$ defined as (\ref{Gm}) are generator matrices of $ \mathfrak{p}_1 $ and $\mathfrak{p}_2$ respectively.

\section{Hole-free Symmetric CRT-based Arrays}\label{Hole-free CRT-based Array}

In general, the elements in the coarrays may not be contiguous, i.e., $\mathcal{D}$ (or equivalently $\mathcal{S}$) may contain holes, which cause ambiguities when subspace-based algorithms are applied. 
In this section, we provide conditions for hole-free and contiguous cross-difference and sum coarrays by modifying the CRT arrays (Definition \ref{crt arrays}) under certain restrictions on quotient rings. 
The definition of hole-free symmetric CRT (HSCRT) arrays is given as follows:

\begin{deft}\label{Hole-free CRT array} [Hole-free Symmetric CRT arrays, HSCRT]
Assume the prime decomposition $  \langle \, p \, \rangle =\mathfrak{p}_1\mathfrak{p}_2$ in $\mathbb{Z}[q]$, with $\mathbf{G}_1$ and $\mathbf{G}_2$ being the generator matrices of $\mathfrak{p}_1$ and $\mathfrak{p}_2$ respectively. A hole-free Symmetric CRT array is an extension of CRT array where $\mathbf{x}_1 \in \sigma(\mathbb{Z}[q]) \, / \, 2\sigma(\mathfrak{p}_1)$ and $\mathbf{x}_2 \in \sigma(\mathbb{Z}[q]) \, / \, \sigma(\mathfrak{p}_2)$ and the two subarrays are $\mathbf{G}_1\mathbf{x}_2$ and $\mathbf{G}_1\mathbf{x}_2$ respectively.
\end{deft}

Note that the two subarrays can be rewritten by means of Voronoi cells (Definition \ref{voronoi cell}) as 
$\sigma(\mathfrak{p}_1) \cap \mathcal{V}(\sigma(\mathfrak{p}_1\mathfrak{p}_2))=\sigma(\mathfrak{p}_1) \cap \mathcal{V}(p\Lambda)$ and  $\sigma(\mathfrak{p}_2) \cap \mathcal{V}(2p\Lambda)$ where $\Lambda=\sigma(\mathbb{Z}[q])$ is the algebraic lattice that corresponds to $\mathbb{Z}[q]$ of a quadratic field.
The following proposition exploits the concept of Voronoi cells to guarantee the 'hole-free' property of HSCRT:

\begin{Proposition} [Generating All Lattice Points in $\Lambda \cap \mathcal{V}(p\Lambda) $]
\label{Generating All Lattice Points}
HSCRT can generate at least all lattice points in $\Lambda \cap \mathcal{V}(p\Lambda)$ by using the cross-difference coarray. 
\end{Proposition}

\emph{Proof}:
For simplicity, let us denote the two ideal lattices by $\Lambda_1= \sigma(\mathfrak{p}_1)$ and $\Lambda_2= \sigma(\mathfrak{p}_2)$ respectively. The ideal is to find a new range for $\mathbf{x}_1$ such that the difference vectors can overspread $\Lambda \cap \mathcal{V}(p\Lambda)$ which corresponds to $\Lambda / p\Lambda$ from quotient group point of view. According to CRT, for all $\mathbf{d} \in \Lambda \cap \mathcal{V}(p\Lambda)$, there exist $\mathbf{x}'_1 \in \Lambda \cap \mathcal{V}(\Lambda_1)$ and $\mathbf{x}_2 \in \Lambda \cap \mathcal{V}(\Lambda_2)$ such that
\begin{eqnarray*}
  \mathbf{d} &\equiv& \mathbf{G}_2 \mathbf{x}'_1 - \mathbf{G}_1 \mathbf{x}_2 \pmod{p\Lambda} \\
   &=& \mathbf{G}_2 \mathbf{x}'_1 - \mathbf{G}_1 \mathbf{x}_2 - \mathbf{y}, \quad \mathbf{y} \in p\Lambda \cap \mathcal{V}(2p\Lambda) \\
   &=& \mathbf{G}_2 \mathbf{x}_1 - \mathbf{G}_1 \mathbf{x}_2, \quad \mathbf{x}_1 = \mathbf{x}'_1 - \mathbf{G}_2^{-1} \mathbf{y}.
\end{eqnarray*}
Considering $\langle \, p \, \rangle = \mathfrak{p}_1 \mathfrak{p}_2$ and their corresponding matrices $\mathbf{G}_1$ and $\mathbf{G}_2$, 
$\mathbf{G}_2^{-1} \mathbf{y}$ is in $\Lambda_1 \, \cap \, \mathcal{V}(2\Lambda_1)$.
Note that $\Lambda_1 $ is a sublattice of $\Lambda$ and $\mathbf{x}'_1 - \mathbf{G}_2^{-1} \mathbf{y}$ is identical to $\mathbf{x}'_1 + \mathbf{G}_2^{-1} \mathbf{y}$ because of the symmetry of $\Lambda_1 $. The proof is completed by noting that $\mathbf{x}_1 \in \Lambda \, \cap \, \mathcal{V}(2\Lambda_1)$. 
In short, by selecting $\mathbf{x}_1 \in \Lambda \, \cap \,  \mathcal{V}(2\Lambda_1)$ and $\mathbf{x}_2 \in \Lambda \, \cap \,  \mathcal{V}(\Lambda_2) $ results in $\mathbf{d} = \Lambda \, \cap \,  \mathcal{V}(p\Lambda)$.

\hfill$\blacksquare$

Generally, the contiguous coarray can be defined as elements within a convex polygon, whereas in this paper, we only consider convex regular polygons such as square and hexagon. 
One remarkable advantage of HSCRT arrays is that because of the symmetry of the algebraic lattices, their cross-difference coarrays are identical to the corresponding sum coarrays. As a result, Proposition \ref{Generating All Lattice Points} also applies to the sum coarrays, implying that both passive and active sensing algorithms that require contiguous coarrays can employ HSCRT arrays.    
The design procedure of HSCRT is summarized in Table \ref{hole free table}.     
Next, we study the properties of HSCRT and formulate the contiguous coarrays of hole-free $\mathbb{Z}^2$ and hole-free $A_2$, which are in the class of HSCRT.

\subsection{Properties of HSCRT Arrays}

\subsubsection{Number of Physical Sensors}
According to Proposition \ref{CRT array}, the number of sensors in $\Lambda_1/p\Lambda$ and in $\Lambda_2/p\Lambda$ both equal to $p$. After doubling the range of $\mathbf{x}_1$ to $ \Lambda \, \cap \,  \mathcal{V}(2\Lambda_1)$ and removing the duplicated sensors at the origin, the total sensor number in $2\Lambda_1$ becomes $4(p-1)$. Thus the number of physical sensors of HSCRT is
\begin{equation*}\label{5p-4}
4(p-1)+p=5p-4. 
\end{equation*}

\subsubsection{Perimeters and Areas of Physical Arrays} Given a prime $p$, the perimeters of hole-free $\mathbb{Z}^2$ denoted as $C_G$ and of hole-free $A_2$ denoted as $C_E$ can be calculated as 
\begin{equation*}
C_G=8pd, \qquad C_E= 6pd(\sin \frac{\pi}{3})^{-1} \approx 6.928 pd,
\end{equation*}
where $d$ is the minimum inter-element spacing and the areas acquired by the two array configurations are
\begin{equation*}
A_G=4p^2d^2,  A_E=3p^2d^2(\sin \frac{\pi}{3})^{-1} \approx 3.464 p^2d^2.
\end{equation*}
Therefore the perimeter and the area of $A_2$ array are about $86\%$ of those $\mathbb{Z}^2$  array, which implies that hole-free $A_2$ is more compact regarding the geometry. 

\subsubsection{Number of Virtual Sensors in Contiguous Coarrays}
According to Proposition \ref{Generating All Lattice Points}, the cross-difference/sum coarray of HSCRT can generate all lattice points in $\Lambda \cap \mathcal{V}(p\Lambda)$, which corresponds to $\Lambda/ p\Lambda$ from the quotient group of view. Since $\Lambda$ is a lattice with generator matrix $\mathbf{G}$, $p\Lambda$ is also a lattice whose generator matrix is $\mathbf{G}\mathbf{B}_p $ where $\mathbf{B}_p = p\mathbf{I}$. The cardinality of $\Lambda \cap \mathcal{V}(p\Lambda)$ equals to the cardinality of $\Lambda/ p\Lambda$ \cite[Definition 3.12.]{oggier2004algebraic}:     
\begin{equation*}
    |\Lambda/ p\Lambda|=|\det(\mathbf{B}_p )|=p^2,
\end{equation*}
i.e., the number of sensors in $\Lambda \cap \mathcal{V}(p\Lambda)$ is $p^2$.

\subsection{Examples of contiguous coarrays of HSCRT} 
\subsubsection{Hole-free $\mathbb{Z}^2$}
\begin{figure}[tb]
  \begin{center}
      \subfigure[]{\includegraphics[height=2.2in]{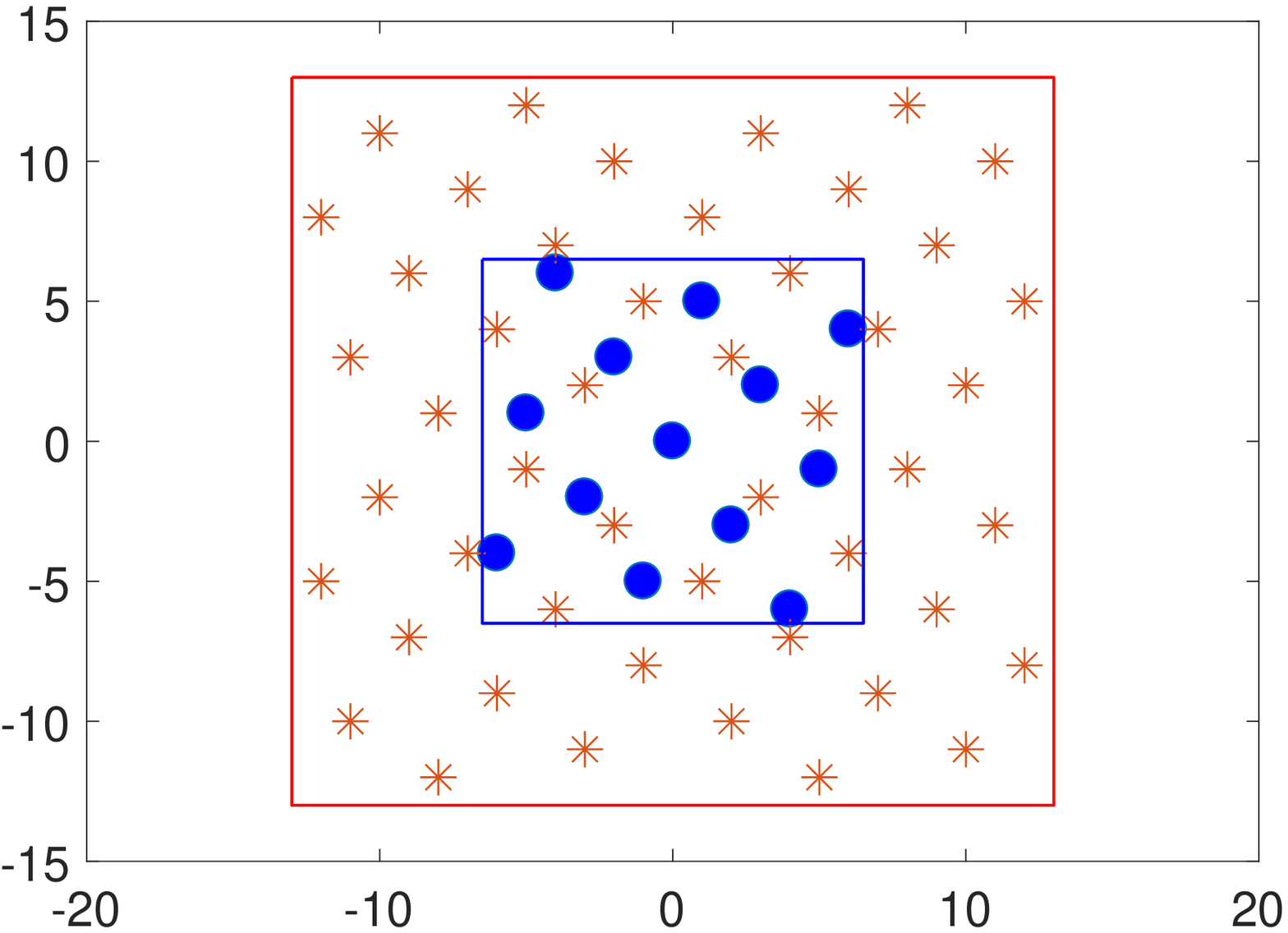}\label{g13fullpositions}}
      \subfigure[]{\includegraphics[height=2.2in]{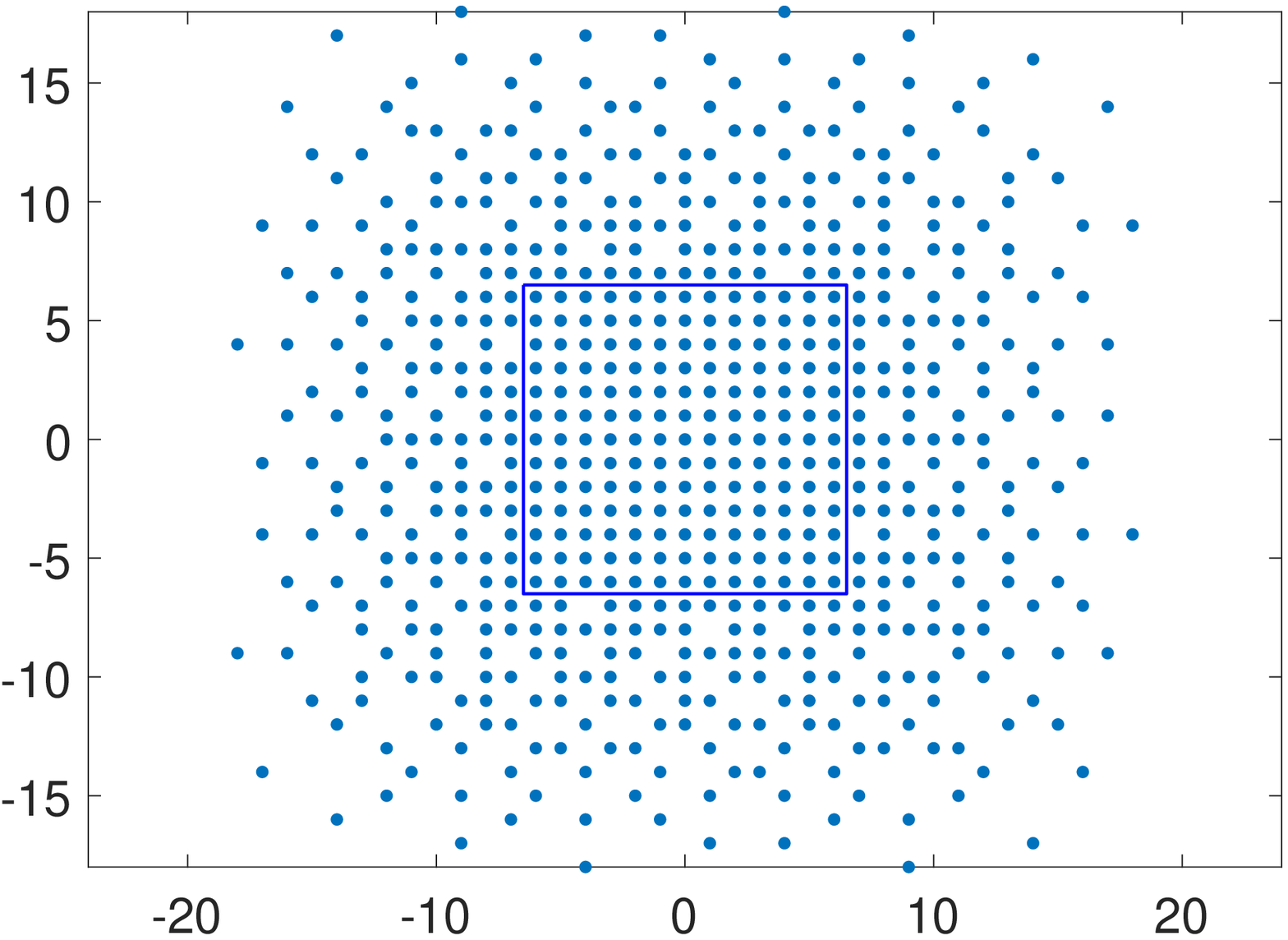}\label{g13fulldifferencecoset1}}
    \caption{Given the decomposition of $p=13$, hole-free $\mathbb{Z}^2$ array (a) and its contiguous cross-difference/sum coarray is within $13\mathbb{Z}^2$ (b). The first subarray $\mathbf{G}_2\mathbf{x}_1$ is in red stars and the second subarray $\mathbf{G}_1\mathbf{x}_2$ is in blue dots. Voronoi cells of $13\mathbb{Z}^2$ and $26\mathbb{Z}^2$ are also shown.}
    \label{g7}
  \end{center}
\vspace{-2em}
\end{figure}

A hole-free $\mathbb{Z}^2$ array is an HSCRT over the ring of Gaussian integers $\mathbb{Z}[i]$, i.e., $\mathfrak{p}_1, \mathfrak{p}_2 \in \mathbb{Z}[i]$ and $\Lambda = \mathbb{Z}^2$. 
The consecutive set of hole-free $\mathbb{Z}^2$ is a uniform rectangular array (URA) which can be expressed as $\mathcal{D}_{C,G}=\mathbb{Z}^2 \cap \mathcal{V}( p\mathbb{Z}^2)$, or equivalently 
\begin{equation*}
\begin{aligned} 
      & \mathcal{D}_{C,G} = \{ \mathbf{d}=(x_k , y_j)  \: | \: \mathbf{d} \in \mathcal{D}, \: -l_G \leq x_k \leq l_G,\\
      & -l_G \leq y_j \leq l_G , k,j = 1, 2, \cdots  l_G \},\\
\end{aligned}
\end{equation*}
where $l_G = \frac{1}{2}p$ according to Proposition \ref{Generating All Lattice Points}. 
It can be verified that the cardinality of $\mathcal{D}_{C,G}$ is $p^2$.
An example of hole-free $\mathbb{Z}^2$ array is depicted in Fig. \ref{g13fullpositions} corresponding to Fig. \ref{g13holepositions} and the effect of filling the holes is illustrated in Fig. \ref{g13fulldifferencecoset1}. 
From this point of view, \cite[Theorem 2]{vaidyanathan2011theory} can be interpreted as a particular case of $\mathbb{Z}^2$.

\begin{figure}[tb]
  \begin{center}
      \subfigure[]{\includegraphics[height=2.2in]{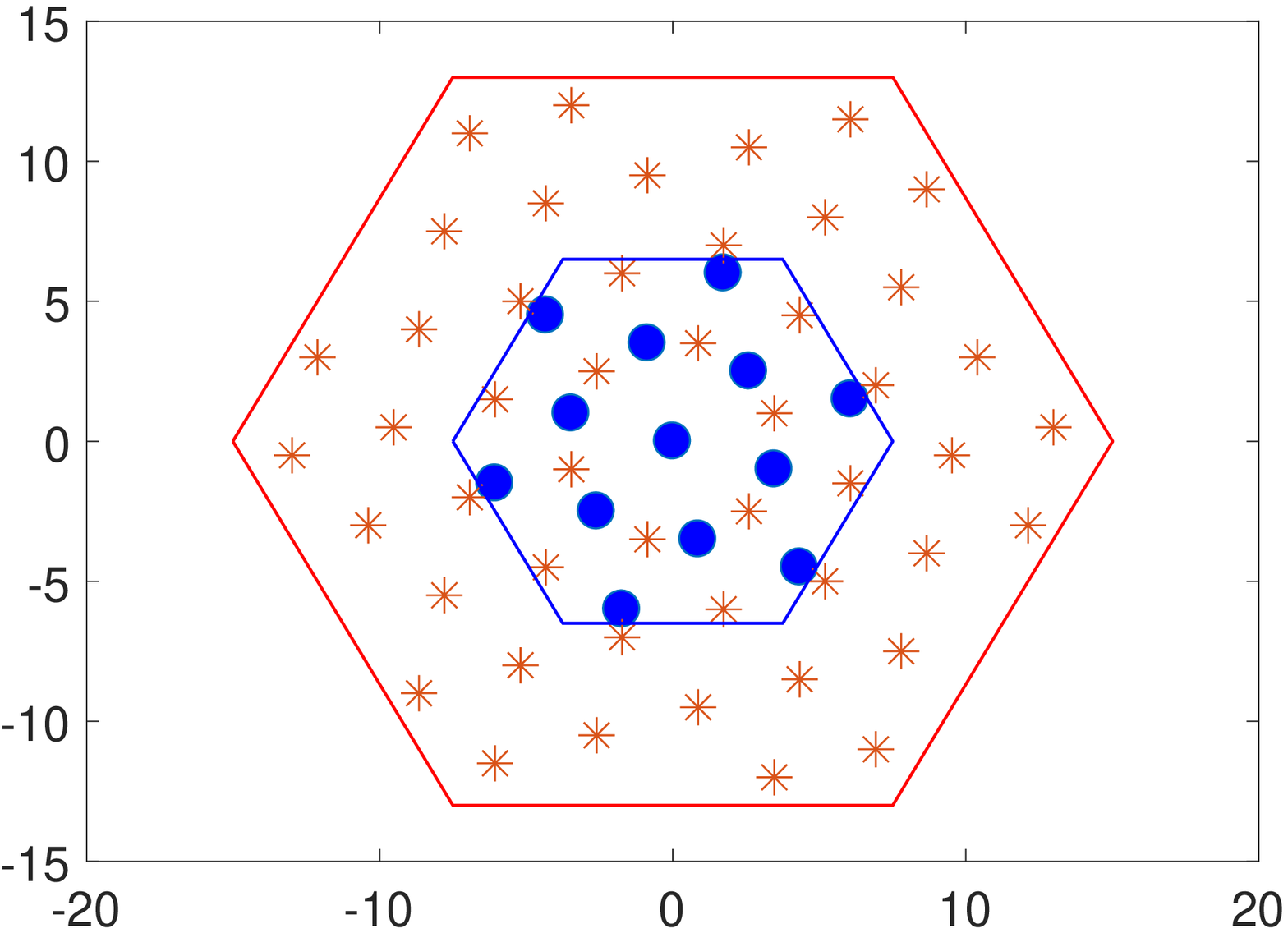}\label{e13fullpositions}}
      \subfigure[]{\includegraphics[height=2.2in]{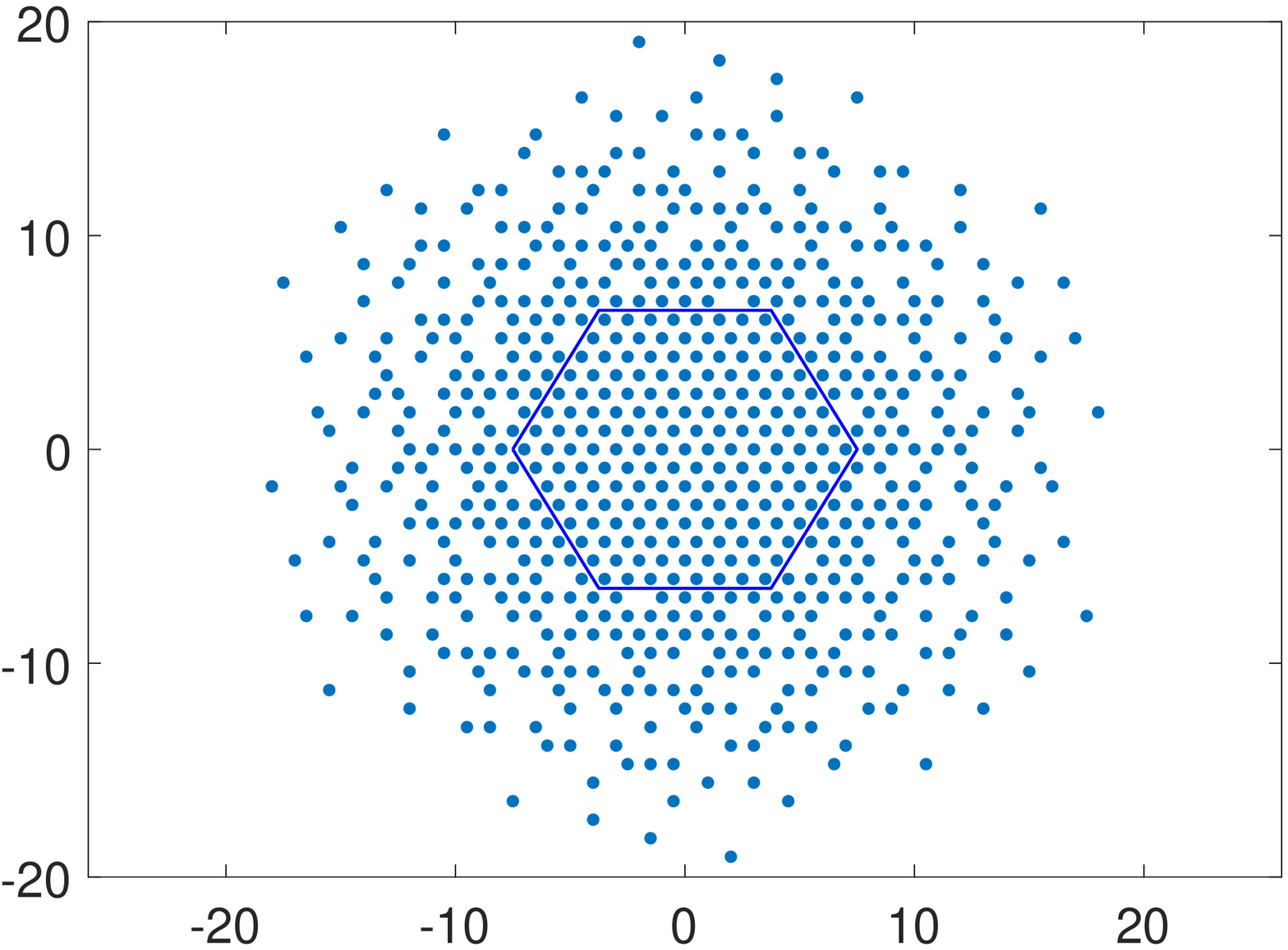}\label{e13fulldifferencecoset1}}
    \caption{Given the decomposition of $p=13$, hole-free $A_2$ array (a) with contiguous cross-difference/sum coarray within $13A_2$ (b). Voronoi cells of $13A_2$ and $26A_2$ are also shown.}
    \label{e7}
  \end{center}
\vspace{-5mm}
\end{figure}

\subsubsection{Hole-free $A_2$}
Analogously, hole-free $A_2$ is defined as a type of HSCRT over the ring of Eisenstein integers $\mathbb{Z}[\omega]$, whose contiguous part of the coarray is also hexagonal with basis given by (\ref{Gee}).
Let $l_r$ denote the inscribed radius of the contiguous hexagonal cell $\mathcal{V}(pA_2)$. Using Proposition \ref{Generating All Lattice Points} and the geometry property of hexagonal lattices, it can be easily verified that $l_r = \frac{1}{2}p$. The contiguous part of the cross-difference/sum coarray of hole-free $A_2$ can be described as $A_2\cap \mathcal{V}(pA_2)$, or equivalently  
\begin{equation*}\label{dce}
\begin{aligned} 
      & \mathcal{D}_{C,E} = \{ \mathbf{d}=(x_k, y_j) \: | \: \mathbf{d}  \in \mathcal{D}, \: - l_r \leq  y_j \leq l_r, \\
      &-2l_r \leq  \pm \sqrt{3} x_k + y_j \leq 2l_r \}
\end{aligned}
\end{equation*}
where there are $ p^2 $ elements in $\mathcal{D}_{C,E}$.
Fig. \ref{e13fullpositions} depicts an example of hole-free $A_2$ with $p=13$ over Eisenstein integers whose cross-difference/sum coarray is shown in Fig. \ref{e13fulldifferencecoset1}.

\section{Conclusion}\label{Conclusion}

In this paper, it has been demonstrated that the problem of designing planar coprime arrays    
can be solved through Chinese remaindering over quadratic fields. 
Inspired by the bijective mappings between the rings of integers and lattices, a new class of array configurations based on coprime lattices constructed from quadratic integers in PIDs is proposed, which provides enhanced DOF and sparse array geometries, and thus alleviates the mutual coupling effect.    
By exploiting the properties of PID, a lattice can be represented by a generator matrix calculated by its corresponding quadratic integer, which significantly simplifies the notations of the sensor locations and generalizes the discussions of coprimality issues. The correlation between coprime quadratic integers and matrices have been investigated in great detail, whereby the coprimality of skew-circulant adjugate pairs can be interpreted as special cases of adjugate matrices in Theorem \ref{adjugate matrices}. 
A modified configuration of CRT array is also introduced for an enlarged contiguous coarray. Examples over $\mathbb{Z}[i]$ and $\mathbb{Z}[\omega]$ are provided for illustrative purposes while in general all quadratic coprime integers can be chosen for CRT array design since the generalized CRT only requires the coprimality of ideals.    

In the accompanying paper, a new approach of obtaining coprime matrices will be demonstrated, after which the multi-sublattice CRT arrays will be introduced, where the subarrays are built from three or more pairwise coprime quadratic integers. The feasibility of the proposed arrays will be employed for both passive and active sensing, which puts forward the algorithms of angle estimations to sparser and more compact hexagonal arrays.

\begin{appendices}

\section{Proof of Theorem \ref{integer matrix}}\label{coprime integer and matrix}
By the assumption on the coprimality of $m$ and $n$, there must exist $\alpha$ and $\beta$ such that $m\alpha + n\beta = 1$ with $m,n,\alpha,\beta \in \mathbb{Z}[q]$. Without loss of generality, let $\{1, q \}$ be an integral basis of $\mathbb{Z}[q]$ where $ q^2 +Bq +C =0 $ according to (\ref{fx2}). Taking the algebraic conjugation of both sides of $m\alpha + n\beta = 1$ yields $\hat{m}\hat{\alpha} + \hat{n}\hat{\beta}  = 1 $ where $\hat{m}$ is the algebraic conjugate of $m$ and same with other elements. $\hat{q}$ is the other root of $f(X)=0$ expressed (\ref{hatq}). By expending all quadratic integers using the basis, it can be easily proved that $m\alpha + n\beta = 1$ if and only if $\hat{m}\hat{\alpha} + \hat{n}\hat{\beta}  = 1 $. Let us write these two equations by the matrix form:
\begin{equation}\label{m n alpha beta P}
    \left(
    \begin{array}{cc}
      m & 0 \\
      0 & \hat{m} \\
    \end{array}
  \right)
      \left(
    \begin{array}{cc}
      \alpha & 0 \\
      0 & \hat{\alpha} \\
    \end{array}
  \right)
  + \left(
    \begin{array}{cc}
      n & 0 \\
      0 & \hat{n} \\
    \end{array}
  \right)    \left(
    \begin{array}{cc}
      \beta & 0 \\
      0 & \hat{\beta} \\
    \end{array}
  \right) \\
  =\mathbf{I}.
\end{equation}
Define $\mathbf{P}_{\alpha}$ and $ \mathbf{P}_{\beta}$ as eigenvalue matrices as $\mathbf{P}_m $ given in (\ref{Pm}). Then (\ref{m n alpha beta P}) can be rewritten as 
$\mathbf{P}_m  \mathbf{P}_{\alpha} + \mathbf{P}_n \mathbf{P}_{\beta} =\mathbf{I}$.
According to Lemma \ref{eigenvectors},  $\mathbf{Q}$ consists the eigenvectors of matrix representations and is expressed in (\ref{Pm}), then left and right multiplying $\mathbf{Q}^{-1}$ and $\mathbf{Q}$ respectively yields
\begin{equation}\label{eq proof bm}
    \mathbf{B}_m  \mathbf{B}_{\alpha} + \mathbf{B}_n \mathbf{B}_{\beta} =\mathbf{I},
\end{equation}
i.e., $\mathbf{B}_m $ and $\mathbf{B}_n $ are left coprime.

Next, we prove the sufficiency of the theorem.
If $\mathbf{B}_m$ and $\mathbf{B}_n$ are assumed to be coprime, there exist $\mathbf{B}_{\alpha}'$ and $\mathbf{B}_{\beta}'$ where 
\begin{equation*}
  \mathbf{B}_{\alpha}'=\begin{pmatrix}
    \alpha'_1 & \alpha'_2\\ \alpha'_3  & \alpha'_4 
  \end{pmatrix}
  \: \, \text{and} \: \,
  \mathbf{B}_{\beta}'=\begin{pmatrix}
     \beta'_1 & \beta'_2\\ \beta'_3  & \beta'_4 
  \end{pmatrix}
\end{equation*}
such that 
$\mathbf{B}_m\mathbf{B}_{\alpha}' + \mathbf{B}_n\mathbf{B}_{\beta}' = \mathbf{I}$, which results the following:
\begin{equation}\label{cop1}
    m_1\alpha'_1 - Cm_2\alpha'_3 + n_1\beta'_1 - Cn_2\beta'_3 = 1, \: \text{and} 
\end{equation}
\begin{equation}\label{cop2}
    m_1\alpha'_3 + m_2\alpha'_1 +n_1\beta'_3+n_2\beta'_1 - Bm_2\alpha'_3 - Bn_2\beta'_3 = 0.
\end{equation}
Let $ \alpha'= \alpha'_1 + \alpha'_3q$ and $ \beta' = \beta'_1 + \beta'_3q $ be two quadratic integers in $\mathbb{Z}[q]$. Then replacing $q^2$ with $-Bq-C$ yields 
\begin{equation}\label{mnalphabelta}
    \begin{aligned}
       & m \alpha' + n \beta' =  (m_1\alpha'_1 - Cm_2\alpha'_3 + n_1\beta'_1 - Cn_2\beta'_3) \\ 
       & + ( m_1\alpha'_3 + m_2\alpha'_1 +n_1\beta'_3+n_2\beta'_1 - Bm_2\alpha'_3 - Bn_2\beta'_3)q
\end{aligned}
\end{equation}
Substituting (\ref{cop1}) and (\ref{cop2}) to (\ref{mnalphabelta}), it can be verified that $ m \alpha' + n \beta' = 1 $.

\section{Proof of Corollary \ref{left right coprime}}\label{appendix left right coprime}

Assume $\mathbf{B}_m$ and $\mathbf{B}_n$ are left coprime. From Theorem \ref{integer matrix}, this assumption is equivalent to that their corresponding quadratic integers $m$ and $n$ in $\mathbb{Z}[q]$ are coprime, i.e., there exist $\alpha$ and $\beta$ such that $m\alpha+n\beta=1$, which is equivalent to (\ref{eq proof bm}). By Lemma \ref{commutative matrices}, (\ref{eq proof bm}) is equivalent to 
\begin{equation*}
     \mathbf{B}_{\alpha}\mathbf{B}_m + \mathbf{B}_{\beta}\mathbf{B}_n=\mathbf{I},
\end{equation*}
i.e., $\mathbf{B}_m$ and $\mathbf{B}_n$ are right coprime.

\section{Proof of Theorem \ref{adjugate matrices}}\label{coprime conditions}

Since the canonical embedding is bijective, the inverse mapping realizes the corresponding algebraic integers of $\mathbf{B}_m$ and $\mathbf{B}_{\hat{m}}$ by $m=m_1+m_2q$ and $\hat{m}=(m_1-Bm_2)-m_2q = m_1 + m_2\hat{q}$ respectively. 
According to Theorem \ref{integer matrix},
Theorem \ref{adjugate matrices} is equivalent to the coprime conditions of two algebraic conjugates in $\mathbb{Z}[q]$. Suppose two conjugate quadratic integers $m$ and $\hat{m}$ are coprime, i.e., $ \text{GCD}( m_1+m_2q, \, m_1 + m_2\hat{q} ) = 1 $. Then $ \text{GCD}( m_1+m_2q, \, m_2( q - \hat{q}) ) = 1 $ by applying Fact (1). This is equivalent to the following two conditions according to Fact (3):
\begin{equation}\label{gcd proof1}
    \text{GCD}( m_1+m_2q, \, m_2 ) = 1, \, \text{and}
\end{equation}
\begin{equation}\label{gcd proof2}
    \text{GCD}( m_1+m_2q, \, \hat{q} - q ) = 1.
\end{equation}
By Fact (1), (\ref{gcd proof1}) can be simplified to 
\begin{equation}\label{adjugate5}
    \text{GCD}( m_1, \, m_2 ) = 1,
\end{equation}
which needs to be held for all cases from (a) to (c). According to (\ref{q}) and (\ref{hatq}), $\hat{q}$ can be replaced by $\hat{q}=-B-q$, then (\ref{gcd proof2}) can be rewritten as 
$ \text{GCD}( m_1+m_2q, \, 1 - 2q) = 1 $ or $ \text{GCD}( m_1+m_2q, \, 2q) = 1 $
corresponding to $B=-1$ or $B=0$ respectively depending on the minimum polynomial of the quadratic field $\mathbb{Q}(\sqrt{D})$ given in (\ref{fx}). 

\begin{enumerate}
\item[(a)] In the first case ($B=-1$), $ \text{GCD}( (2m_1+m_2)q, 1 - 2q) = 1 $ is obtained by subtracting $m_1 ( 1 - 2q ) $ to the first entry, which is equivalent to
\begin{equation*}\label{gcd proof3}
     \text{GCD}( 2m_1 + m_2, \, 1-2q ) = 1, 
\end{equation*}
since $\text{GCD}(q, 1-2q) = \text{GCD}(1,q) = 1 $ (1 and $q$ must be coprime as $\{ 1, q\} $ is an integral basis).
Substituting $B=-1$ to $f(X)$, (\ref{fx2}) becomes $q^2 - q + C = 0 $. Thus (a) is obtained by enforcing a square on the second entry, i.e., $(1-2q)^2 = 1-4C $. Recall (\ref{adjugate5}) shall hold. 

\item[(b)] With $B=0$ (the GCD of $m_1+m_2q$ and $2q$ is 1), enforcing a square on $m_1+m_2q$ and applying Fact (1) result in $\text{GCD}( m_1^2+m_2^2q^2, 2q) = 1$, which is equivalent to $\text{GCD}( m_1^2+m_2^2q^2, q)= 1$ and $\text{GCD}( m_1^2+m_2^2q^2, 2)= 1$ by Fact (3). Note that $q^2=C$ given $B=0$ (\ref{q}). Thus by Fact (1)-(3), the former can be simplified to 
\begin{equation}\label{gcd proof4}
    \text{GCD}( m_1^2, q)=\text{GCD}( m_1, C)= 1,
\end{equation}
and the latter becomes $\text{GCD}( m_1^2+m_2^2C, 2)= 1$, which  
can be simplified depending on the parity of $C$ as follows:  

If $C$ is an even number, $2$ divides $Cm_2^2$ and thus it can be eliminated, resulting in $ \text{GCD}(m_1^2, \, 2)=1$, which is equivalent to 
\begin{equation}\label{gcd proof5}
    \text{GCD}(m_1, \, 2)=1
\end{equation}
by Fact (2), i.e., $m_1$ is odd, which coincides with (\ref{gcd proof4}) with even $C$. Recall (\ref{adjugate5}) shall hold.

\item[(c)] Likewise, when $C$ is odd, $C$ can be viewed as a sum of a even number $C'$ and the unit, i.e., $\text{GCD}(m_1^2+(2C'+1)m_2^2, \, 2)=1$. Therefore $C'm_2^2$ can be eliminated as the previous case, after which Fact (1) and Fact (3) can be applied, i.e.,
\begin{equation}\label{gcd proof6}
    \text{GCD}((m_1+m_2)^2, \, 2)=\text{GCD}(m_1+m_2, \, 2)=1.
\end{equation} 
Recall (\ref{adjugate5}) and (\ref{gcd proof4}) shall hold. By repeatedly applying Fact (1) and Fact (3), (\ref{adjugate5}) and (\ref{gcd proof6}) can be incorporated together since (\ref{adjugate5}) can be rewritten as $\text{GCD}( m_1 + m_2, \, m_2 ) = 1$ and $\text{GCD}(m_1+m_2, \, 2m_2) = \text{GCD}(m_1+m_2, \, m_1+m_2 - 2m_2) $.    
\end{enumerate}

\end{appendices}

\bibliographystyle{ieeetr}
\bibliography{reference}

\end{document}